\documentclass[preprint2,twocolappendix]{aastex6}

\usepackage{longtable}
\usepackage{rotating}

\shorttitle{ALMA AGN Sizes}
\shortauthors{Chang et al.}

\begin{document}


\title{Unveiling Sizes of Compact AGN Hosts with ALMA}



\author{Yu-Yen~Chang\altaffilmark{1,2}}
\email{yuyenchang.astro@gmail.com}
\author{Emeric~Le~Floc'h\altaffilmark{2}}
\author{St\'ephanie~Juneau\altaffilmark{3}}
\author{Elisabete~da~Cunha\altaffilmark{4,5,6}}
\author{Mara~Salvato\altaffilmark{7}}
\author{Avishai Dekel\altaffilmark{8}}
\author{Francesca~Civano\altaffilmark{9}}
\author{Stefano~Marchesi\altaffilmark{10}}
\author{Hyewon Suh\altaffilmark{11}}
\author{Wei-Hao Wang\altaffilmark{1}}
\altaffiltext{1}{Academia Sinica Institute of Astronomy and Astrophysics, PO Box 23-141, Taipei 10617, Taiwan}
\altaffiltext{2}{CEA Saclay, DSM/Irfu/Service d'Astrophysique, Orme des Merisiers, F-91191 Gif-sur-Yvette Cedex, France}
\altaffiltext{3}{National Optical Astronomy Observatory, 950 N Cherry Avenue, Tucson AZ 85719, USA}
\altaffiltext{4}{International Centre for Radio Astronomy Research, University of Western Australia, 35 Stirling Hwy, Crawley, WA 6009, Australia}
\altaffiltext{5}{Research School of Astronomy and Astrophysics, The Australian National University, Canberra, ACT 2611, Australia}
\altaffiltext{6}{ARC Centre of Excellence for All Sky Astrophysics in 3 Dimensions (ASTRO 3D)}
\altaffiltext{7}{Max Planck Institut f\"ur Plasma Physik and Excellence Cluster, D-85748 Garching, Germany}
\altaffiltext{8}{Racah Institute of Physics, The Hebrew University, Jerusalem 91904 Israel}
\altaffiltext{9}{Harvard-Smithsonian Center for Astrophysics, 60 Garden Street, Cambridge, MA 02138, USA}
\altaffiltext{10}{Department of Physics \& Astronomy, Clemson University, Clemson, SC 29634, USA}
\altaffiltext{11}{Subaru Telescope, National Astronomical Observatory of Japan (NAOJ), National Institutes of Natural Sciences (NINS), 650 North A'ohoku place, Hilo, HI, 96720, USA}


\begin{abstract}
We present rest-frame far-infrared (FIR) and optical size measurements of AGN hosts and star-forming galaxies in the COSMOS field, enabled by high-resolution ALMA/1~mm  ($0\arcsec.1$-$0\arcsec.4$) and HST/$F814W$ imaging ($\sim0\arcsec.1$). Our sample includes 27 galaxies at $z<2.5$, classified as infrared-selected AGN (3 sources), X-ray selected AGN (4 sources), and non-AGN star-forming galaxies (20 sources), for which high-resolution Band 6/7 ALMA images are available at 1~mm from our own observing program as well as archival observations. The sizes and SFR surface densities measured from both ALMA/1~mm and HST/$F814W$ images show that obscured AGN host galaxies are more compact than non-AGN star-forming galaxies at similar redshift and stellar mass. This result suggests that the obscured accretion phase may be related to galaxies experiencing a compaction of their gaseous component, which could be associated with enhanced central star formation before a subsequent quenching driving the formation of compact passive galaxies. Moreover, most of the detected and stacked rest-frame FIR sizes of AGNs in our sample are similar or more compact than their rest-frame optical sizes, which is consistent with recent results of ALMA detected sources. This might be explained by the fact that the dusty starbursts take place in the compact regions, and suggests that the star formation mechanisms in the compact regions of AGN hosts are similar to those observed in star-forming galaxies observed with ALMA. 
\end{abstract}





\section{Introduction}
\label{sec1}

The processes driving the co-evolution of galaxies and their super-massive black holes remain a largely debated issue in extragalactic astrophysics. 
Specifically, the connection between the formation of stars in galaxies and the fueling of their central black holes is still not fully understood.
Previous results revealed that X-ray and optically selected Active Galactic Nuclei (AGNs) reside in galaxies harboring sustained activity of star formation \citep[e.g.,][]{2012ApJ...760L..15H,2012MNRAS.419...95M,2013ApJ...764..176J,2013ApJ...771...63R}. 
Furthermore, although high luminosity AGNs seem to be connected to violent events (i.e., quasars) and may reside in  galaxies more often involved in major mergers \citep[e.g.,][]{2012ApJ...757...23K,2009ApJ...696..110T}, the majority of moderate-luminosity X-ray selected AGNs appear to live mostly in disk-dominated isolated systems \citep{2011A&A...535A..80M,2011ApJ...743....2S,2014ApJ...784L...9F,2014MNRAS.439.3342V}. 
Also, major mergers do not appear to be the dominating triggering mechanism for luminous unobscured AGN \citep{2016ApJ...830..156M,2017MNRAS.466..812V}.
Their  morphology  should thus be very similar to that of sources populating the so-called main sequence \citep[stellar mass vs.\ SFR; e.g., ][]{2004MNRAS.351.1151B,2007A&A...468...33E,2007ApJ...660L..43N,2007ApJ...670..156D,2015A&A...579A...2I,2015A&A...575A..74S,2015ApJS..219....8C} of star-forming galaxies \citep[e.g.,][]{2009ApJ...691..705G,2011ApJ...726...57C,2012ApJ...744..148K}, implying that the bulk of super-massive black hole accretion is likely driven by internal processes and not by major mergers. Nonetheless, it is still an open question whether the internal structure of AGN hosts is comparable or different from that of normal star-forming galaxies.

Most current studies have focused on samples of star-forming galaxies.
Recent observations from the Atacama Large Millimeter/submillimeter Array (ALMA) allow us to measure far-infrared (FIR) sizes with high angular resolution and sensitivity \citep[e.g.,][]{2015ApJ...810..133I,2015ApJ...807..128S,2016ApJ...833..103H}.
For size comparison, it has been discussed that rest-frame FIR sizes of massive star-forming galaxies are smaller than their rest-frame optical sizes \citep[e.g., ][]{2015ApJ...799..194C,2016ApJ...827L..32B,2017ApJ...841L..25T,2017ApJ...850...83F,2018ApJ...861....7F,2018A&A...616A.110E,2019ApJ...879...54L}.
A possible explanation of the central compact component is a compact dusty bulge which related to bulge formation in the center \citep{2017ApJ...850...83F}. Moreover, it has been discussed that massive galaxies turn out to exhibit compact star-formation \citep{2016ApJ...827L..32B,2018A&A...616A.110E}, and transform into compact quiescent galaxies at $z\sim2$ \citep[e.g.,][]{2014ApJ...788...28V}.
There have been several attempts to estimate FIR sizes for faint submillimeter galaxies \citep[SMGs][]{2016ApJ...833...12R,2017ApJ...850...83F, 2017A&A...597A..41G}, but the uncertainties are still constrained by the small number statistics. Moreover, it has been discussed that both high sensitivities and high angular resolutions are required to measure FIR sizes for less massive objects \citep{2018ApJ...861....7F}.
On the other hand, it is still unclear whether the existence of AGNs affects the structures of their host galaxies, and whether AGN host galaxies show different rest-frame FIR sizes compared to other star-forming galaxies.

\citet{2017MNRAS.466L.103C} investigated AGN and non-AGN host morphologies through HST/$F814W$ images, and found that obscured IR-selected AGN hosts are more compact than normal star-forming sources at a given stellar mass at $z\sim1$.
Following that work, in this paper we resolve the dust emission of AGN host galaxies using high-resolution ALMA imaging. We show that their compact sizes are consistent with what we found using HST/$F814W$-band imaging, and the FIR (ALMA/~1 mm) sizes are yet even smaller than the optical sizes.
We compare all available FIR sizes of AGN and non-AGN hosts in the Cosmic Evolution Survey \citep[COSMOS;][]{2007ApJS..172....1S} field.
For comparison, we also measure the FIR sizes of a control sample of non-AGN star-forming galaxies in the COSMOS field for which we find complementary high-resolution ALMA imaging in the archive. 
The structure of this paper is as follows.
We describe the data and sample selections in Section \ref{sec2}.
We analyze the physical and structural properties in Section \ref{sec3}.
We discuss the results in Section \ref{sec4} and summarize in Section \ref{sec5}.
Throughout the paper, we use AB magnitudes, adopt the cosmological parameters
($\Omega_{\rm M}$,$\Omega_\Lambda$,$h$)=(0.30,0.70,0.70), and assume the
stellar initial mass function of \citet{2003PASP..115..763C}.


\begin{figure*}
\centering
\includegraphics[width=2.0\columnwidth]{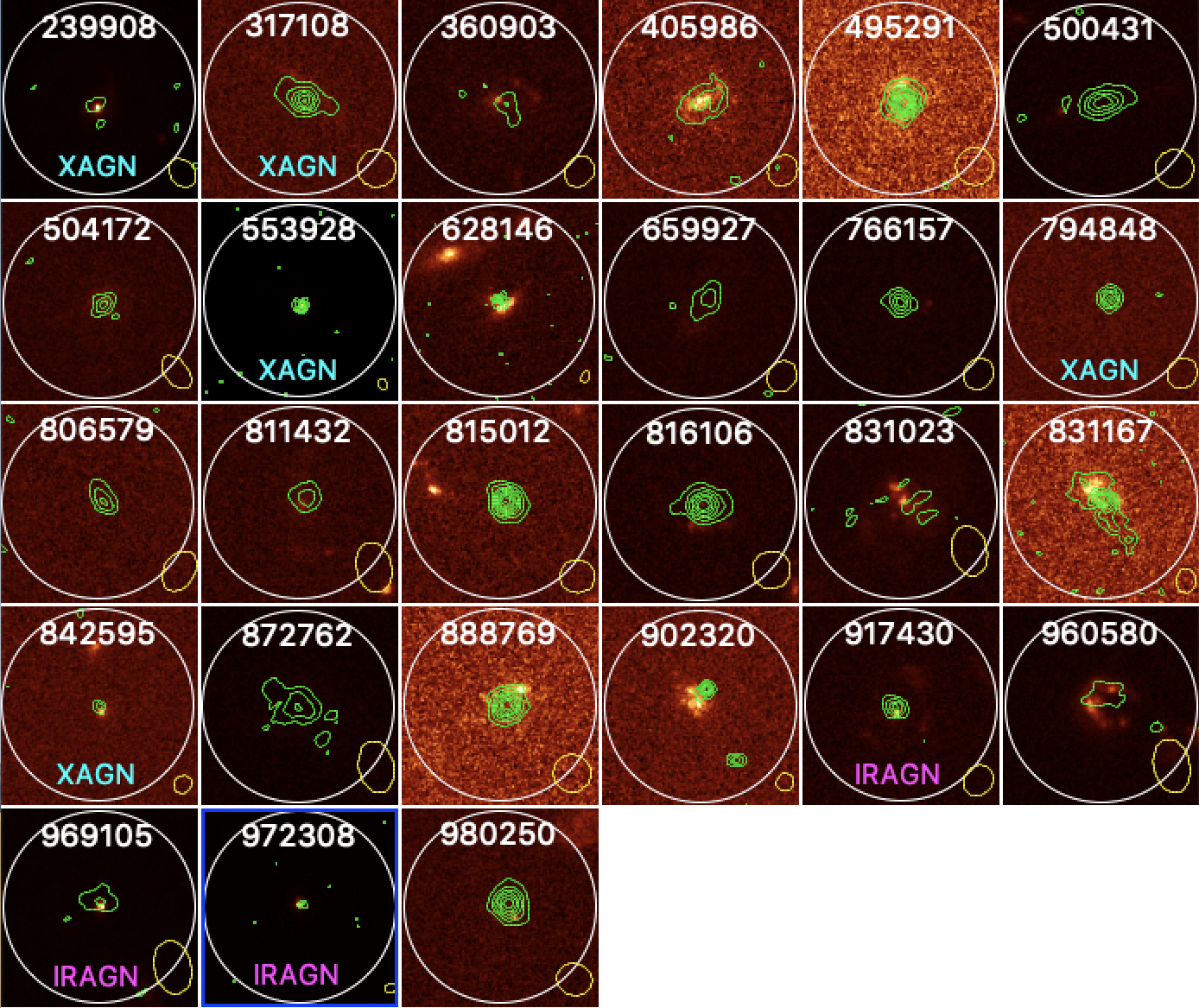} 
\caption{HST/$F814W$ band images for the 27 detected sources. The green contours show the ALMA band 6/7 intensities from the 3 to 24 $\sigma$ with a 3 $\sigma$ step. The yellow circles represent the beam size. The white circles are the 2$\arcsec$ radii of the optical objects. The numbers are the COSMOS2015 IDs. The infrared-selected AGNs are labeled as IRAGN, and the X-ray selected AGNs are labeled as XAGN. Our detected source (972308) is in the blue frame.}
\label{alma_img}
\end{figure*}

\begin{figure}
\centering
\includegraphics[width=0.6\columnwidth]{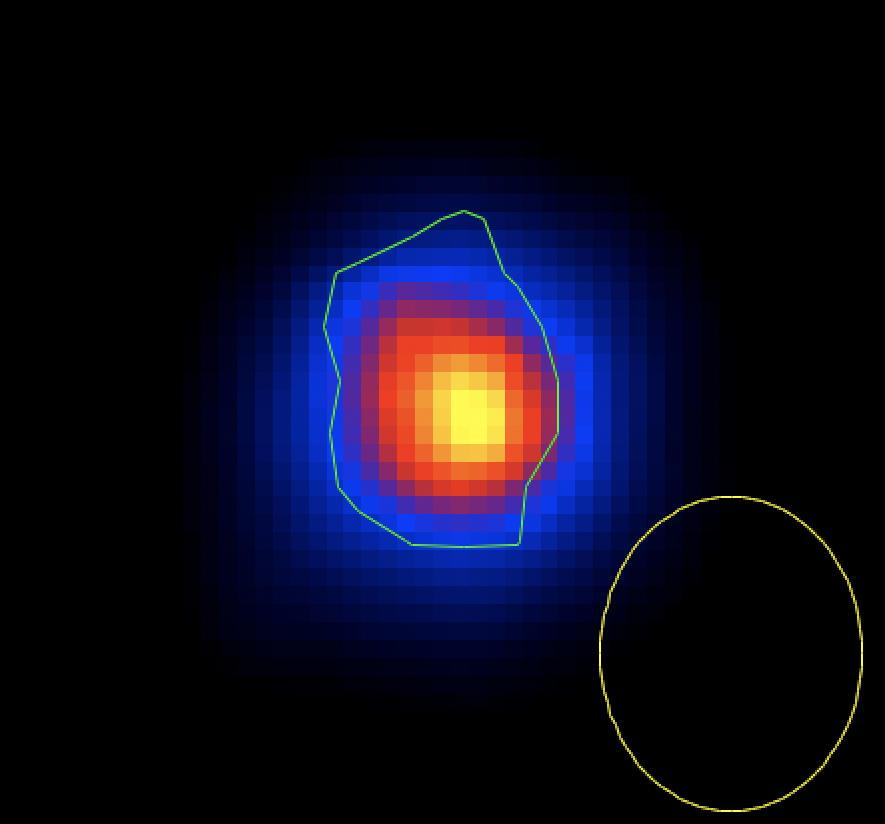} 
\caption{Stacked image of the 9 undetected sources. The background is the HST/$F814W$ image, and the green contour shows the ALMA/1~mm  intensities of the 3 $\sigma$ detection. The yellow circle represents the beam size. The box size is 5 kpc$\times$ 5 kpc. }
\label{alma_stack}
\end{figure}


\section{Samples and Observations}
\label{sec2}

\subsection{Our ALMA Observations}
Our ALMA Band 7 (275-373GHz) observations (\#2016.1.01355.S) contain ten obscured AGN host galaxies selected in the infrared \citep[based on identifying mid-IR power-law signatures which are characteristic of AGNs as opposed to star-forming dominated sources; more details in][]{2017ApJS..233...19C} at 0.5$<$z$<$1.0 in the COSMOS field.
Our targets were selected along the main sequence within 0.3 dex of star formation, as well as star-forming sample in the rest-frame $UVJ$ plot \citep{2009ApJ...691.1879W}, with stellar masses $\log \rm{M}_* >10.5$ in a 24$\mu$m selected sample in the COSMOS field. Reliable spectroscopic redshifts are available.
In order to ensure a robust estimate of the dust-obscured star formation, we restricted our sample to \emph{Herschel}-detected sources. We confirmed that these objects show compact $F814W$-band radial profiles as described in \citet{2017MNRAS.466L.103C}. 

We proposed the standard set-up for single continuum centered at 343.5 GHz of Band 7 over a 7.5 GHz bandwidth.
The proposed angular resolution is 0$\arcsec$.1, which is close to the resolution of the $F814W$-band imaging. 
The array configuration was C40-6 with 40 antennas, and the proposed rms value is 0.1 mJy/beam per 7.5 GHz. 
The data were reduced with the Common Astronomy Software Applications package \citep[CASA; ][]{2007ASPC..376..127M}, and the steps of data reduction and calibration were performed with the scripts provided by the ALMA observatory.
The entire map was produced by the CLEAN algorithm with the \texttt{TCLEAN} task.
The final images are characterized by a synthesized beam size of 0$\arcsec$.12 $\times$  0$\arcsec$.10 \footnote{The synthesized beam sizes are slightly different if we use different weighting.  A synthesized beam size of 0$\arcsec$.12 $\times$  0$\arcsec$.10 corresponds to Briggs weighitng with 0.5 robustness parameter in CASA.} as shown in Table~\ref{tab_0}.
One of the objects (Source name: 972308) is detected, and is shown in Figure~\ref{alma_img} along with the remaining detections from the ALMA archive. The rms value is 0.096 mJy/beam per 8 GHz bandwidth ($\sim$ 0.099 mJy/beam per 7.5 GHz). For the remaining nine objects, we stacked their imaging and computed the median values in each pixel to create a stacked imaging as shown in Figure~\ref{alma_stack}. The peak flux is 0.09 mJy$\pm$0.03 mJy (3.39 $\sigma$; rms value is calculated from the stacked image) \footnote{In order to have high signal-to-noise ratio, we used natural weighting with \texttt{TCLEAN} in CASA. As a result, the synthesized beam size becomes 0$\arcsec$.14 $\times$  0$\arcsec$.12.}.  
We also stacked these objects by \texttt{STACKER} \citep{2015MNRAS.446.3502L} in the uv plane. 
The peak flux is 0.08 mJy$\pm$0.03mJy (2.24 $\sigma$; rms value is calculated from the output image of \texttt{STACKER}), which is consistent with the stacking results from the image plane. 
For point sources, nine objects with a rms of 0.1 mJy/beam can produce at best a stacked image with a rms of 0.03 mJy/beam, and given that the stacked signal is 0.09 the signal-to-noise ratio would be around 3. Our stacking result from image plane shows slightly large signal-to-noise ratios, which could be explained by the extended profile and the beam size.
It is not straightforward to convert stacked image in the uv plane from angular scale to physical scale, so we adopted the FIR stacked image in the image plane in this paper.

\subsection{Complementary Sample from the Archive}
We searched all available ALMA public data by 2019 May to acquire a complementary sample. 
We found 27 ALMA detected sources as shown in Figure~\ref{alma_img}, including AGNs and non-AGN galaxies in Band 6/7 with angular resolution smaller than 0$\arcsec$.4 and peak flux values larger than 3 $\sigma$ at $z<2.5$, and the one in our program.
Table~\ref{tab_0} shows the properties of these archival data (\#2013.1.00034.S, \#2015.1.00026.S, \#2016.1.00478.S, \#2016.1.00735.S, \#2016.1.00778.S, \#2016.1.01355.S, \#2016.1.01426.S, and \#2016.1.01604.S.). 
The angular resolution is in the range of $0\arcsec.1$-$0\arcsec.4$.
We simulated ALMA images by assuming Gaussian profiles, observed the images by \texttt{SIMOBSERVE}, and analyzed the observations by \texttt{SIMANALYZE} with similar inputs as our program (single continuum centered at 343.5 GHz of Band 7 over a 7.5 GHz bandwidth) in CASA. Then we used the \texttt{UVMODELFIT} task to fit single component Gaussian models to the UV data, and calculated the fitted sizes. We assumed sizes in a range of $0\arcsec.10$ to $0\arcsec.40$ (0.8 kpc to 3.2 kpc at z=1; 0.9 kpc to 3.3 kpc at z=2), as well as fluxes in a range of 0.25 mJy to 2.00 mJy (signal-to-noise ratios in a range of 4 to 50 with 300 seconds total time). For high resolution observations (0$\arcsec$.1-0$\arcsec$.4) close to our program, the size measurements are increasing from underestimation to overestimation. The underestimation can be caused by over-resolving low signal-to-noise ratio objects, and the overestimation can be caused by insufficient resolutions. Nevertheless, the measured size differences are within 20\% of the simulated galaxies if the resolution is around $0\arcsec.1$ to $0\arcsec.4$. For spatial resolution larger than 0$\arcsec$.5, the size measurements from \texttt{UVMODELFIT} are getting larger and overestimated by more than 30\% size differences, and the profiles sometimes fail to be fitted, especially for small sources or low signal-to-noise ratio objects.
We also did a simulation by \texttt{IMSMOOTH} function in CASA to create images with $0\arcsec.4$ angular resolution for images with $0\arcsec.1$ angular resolution. Most sources ($>$90\%) are well-fitted and their differences of sizes are within 10\%, so we confirmed that these data can be complementary to our program \footnote{For a choice of $1\arcsec.0$ ($0\arcsec.8$) simulated images, the differences of sizes are 40\% (30\%). Moreover, $>$90\% sources in the range of $0\arcsec.8$-$1\arcsec.0$ fail to be well-fitted in the image plane, and $>$50\% sources in the range of $0\arcsec.4$-$0\arcsec.8$ fail to be well-fitted in the uv plane. Considering the simulation and uncertainties of the fitting results, we decided to have a range of $0\arcsec.1$-$0\arcsec.4$ as our complementary sample.}.
We matched these sources with the COSMOS2015 catalog \citep{2016ApJS..224...24L}.
The redshifts we adopted are firstly taken from the Chandra COSMOS Legacy Survey \citep[14 sources;][]{2011ApJ...742...61S,2016ApJ...817...34M,2016ApJ...819...62C}, then from the COSMOS2015 photometric redshifts \citep[13 sources;][]{2016ApJS..224...24L}.
Among the 27 sources as shown in Table~\ref{tab_0}, 13 of them have reliable spectroscopic redshifts, 23 of them are detected by \emph{Herschel} ($S/N>3$ in any \emph{Herschel} bands), 3 of them are X-ray selected AGNs, and 3 of them are infrared selected AGNs according to \citet{2017ApJS..233...19C}.

\subsection{SED Fitting Results}
We fitted the spectral energy distributions (SEDs) of these 27 sources from the UV to infrared with an updated version of the MAGPHYS fitting technique \citep{2008MNRAS.388.1595D,2015ApJ...806..110D}. The version we used accounts for a possible AGN contribution assuming empirical AGN components from a set of empirical templates as described in \citet{2017ApJS..233...19C}: \citet[type-2]{2011MNRAS.414.1082M}, \citet[QSO]{2006ApJS..166..470R}, \citet[QSO]{,2010MNRAS.402..724P}, and \citet[Seyfert 1]{2007ApJ...663...81P}. 
We included ALMA detections or upper limits in addition to the photometry in \citet{2017ApJS..233...19C}.
The stellar masses of these 27 sources are all above $\log \rm{M}_* =10.5$, which is complementary to our selected sample. 
It has been known that the energy balance for UV and FIR in SED can break down because the dominant optical and IR emission of these galaxies do not come from the same physical regions \citep[see][ for a recent example]{2017ApJ...844L..10S}.
Therefore, we calculated the decomposed infrared and UV luminosity from the star formation component after subtracting the AGN contribution by our SEDs. Then we calculated the SFRs by the sum of these decomposed infrared and UV SFRs according to \citet{1998ApJ...498..541K,2012ARA&A..50..531K}. 
We found that the SFRs calculated by this method are about 0.17$\pm$0.02 dex higher than the original SFRs estimated by SED fitting. However, estimating sizes and estimating SFR are two different things that do not impact each other, because the SFR differences for IRAGN (0.20$\pm$0.06 dex), XAGN (0.16$\pm$0.04 dex), and the SF (0.19$\pm$0.03 dex) galaxies are very close.
We also visually inspected the SED fitting results, and found that the fit quality was adequate (i.e., no major failures). As a result, we adopt the estimation above in this paper. 

\subsection{Size Measurements}
The effective radius in $F814W$-band is about 3kpc ($\sim$0$\arcsec$.4 at $z=0.75$) and the FWHM of the HST/$F814W$ imaging is 0$\arcsec$.09, so high resolution ALMA observations ($<$0$\arcsec$.4) are required to resolve the dust distribution at a complementary angular resolution of the HST imaging.

First, we measured the rest-frame FIR sizes with the \texttt{IMFIT} task in the image plane, which fitted two-dimensional elliptical Gaussian components, in CASA (version 5.1.0). The effective radius by \texttt{IMFIT} is calculated by $R_e=\sqrt{R_{Maj} \times R_{Min}}/2$, where $R_{Maj}$ and $R_{Min}$ are the deconvolved FWHM major and minor axis. 
Then we used the \texttt{UVMODELFIT} task in the uv plane, which fits a single component source model to the UV data. We used the flux density and the size measurements obtained by the \texttt{IMFIT} task as the initial values, and fitted Gaussian models with five iterations by \texttt{UVMODELFIT} task. The effective radius by \texttt{UVMODELFIT} is also calculated by the fitted major and minor axis.
We compared the measurements by \texttt{IMFIT} for the image plane and \texttt{UVMODELFIT} for the uv plane. The difference of the sizes are with a 2\% systematic difference (\texttt{UVMODELFIT} gives lower values).
To avoid the question about the fidelity of the ALMA imaging, we adopted the size measurements from \texttt{UVMODELFIT} for single objects in this paper.
We also measured the rest-frame FIR sizes with \texttt{GALFIT} \citep{2010AJ....139.2097P}, assuming galaxy models with single S\'ersic profiles. The sizes measured from \texttt{GALFIT} are slightly larger (systematic difference $\sim$10\%) than the sizes measured from \texttt{IMFIT}. These comparisons show that our size measurements are consistent with each other.

There are 23 out of 27 sources with available HST/$F814W$ images. We measured the $F814W$-band light distribution of our COSMOS galaxy sample with \texttt{GALFIT}, assuming galaxy models with the single S\'ersic profiles. 
Among the 27 sources with reliable redshifts, 2 of them can be found in the Cosmic Assembly Near-infrared Deep Extragalactic Legacy Survey (CANDELS, \citealp{2011ApJS..197...35G,2011ApJS..197...36K}) region. Therefore, we used the single S\'ersic profile measurement of available $F125W$ and $F160W$ imaging from \citet{2012ApJS..203...24V} in the CANDELS/COSMOS \citep{2011ApJS..197...35G,2011ApJS..197...36K} field. The effective radius from single S\'ersic profile represents the rest-frame optical size.


\begin{figure*}
\centering
\includegraphics[width=1.75\columnwidth]{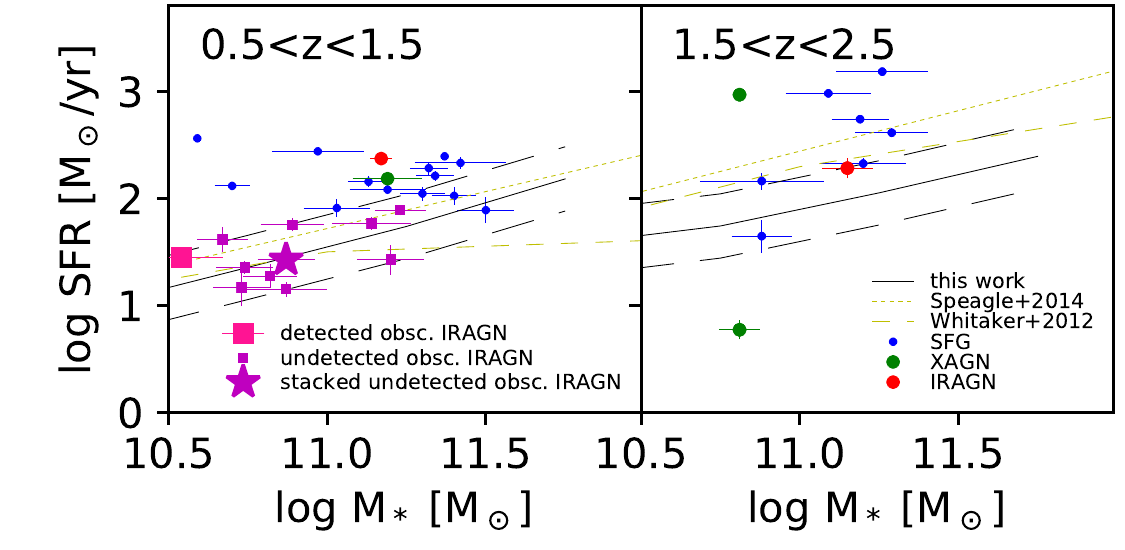} 
\caption{SFR versus stellar mass plots at $z\sim1$ and $z\sim2$. The black solid and dashed lines are the star-forming sequence defined by our SED fitting results. 
The detected obscured IRAGN (pink square), undetected obscured IRAGN (purple square), and stacked values of the undetected obscured IRAGNs (star) in our ALMA sample are shown. The blue dots are star-forming galaxies (SFG), green dots are X-ray selected AGN hosts (XAGN), and red dots are infrared selected AGN hosts (IRAGN) from the ALMA archive.   Most of our obscured IRAGNs have lower stellar masses and SFRs. The main sequence derived by \citet{2014ApJS..214...15S} (yellow dotted lines)  and \citet{2014ApJ...795..104W} (yellow dash lines) are also shown.}
\label{alma_msfr}
\end{figure*}

\begin{figure*}
\centering
\includegraphics[width=2.00\columnwidth]{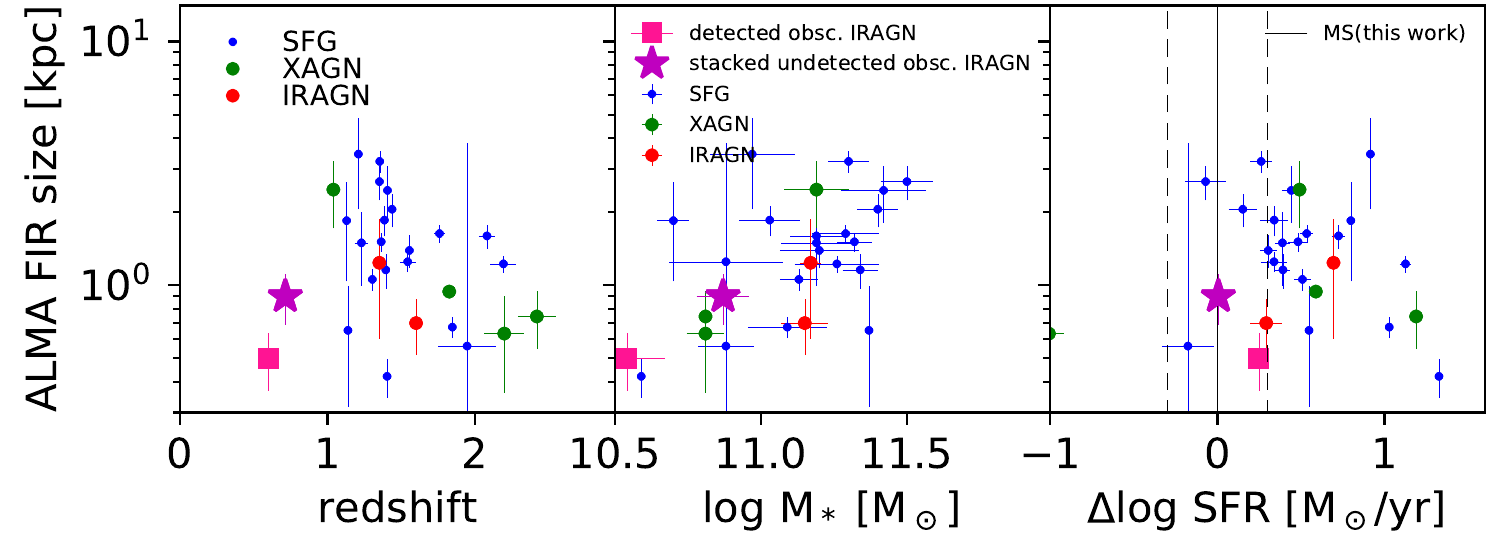} 
\caption{Rest-frame FIR size (effective radius) versus redshift, stellar mass , and distance to the star-forming sequence according to our SED fitting results. The symbols are the same as described in Figure~\ref{alma_msfr}.}
\label{alma_rzm}
\end{figure*}

\begin{figure*}
\centering
\includegraphics[width=1.75\columnwidth]{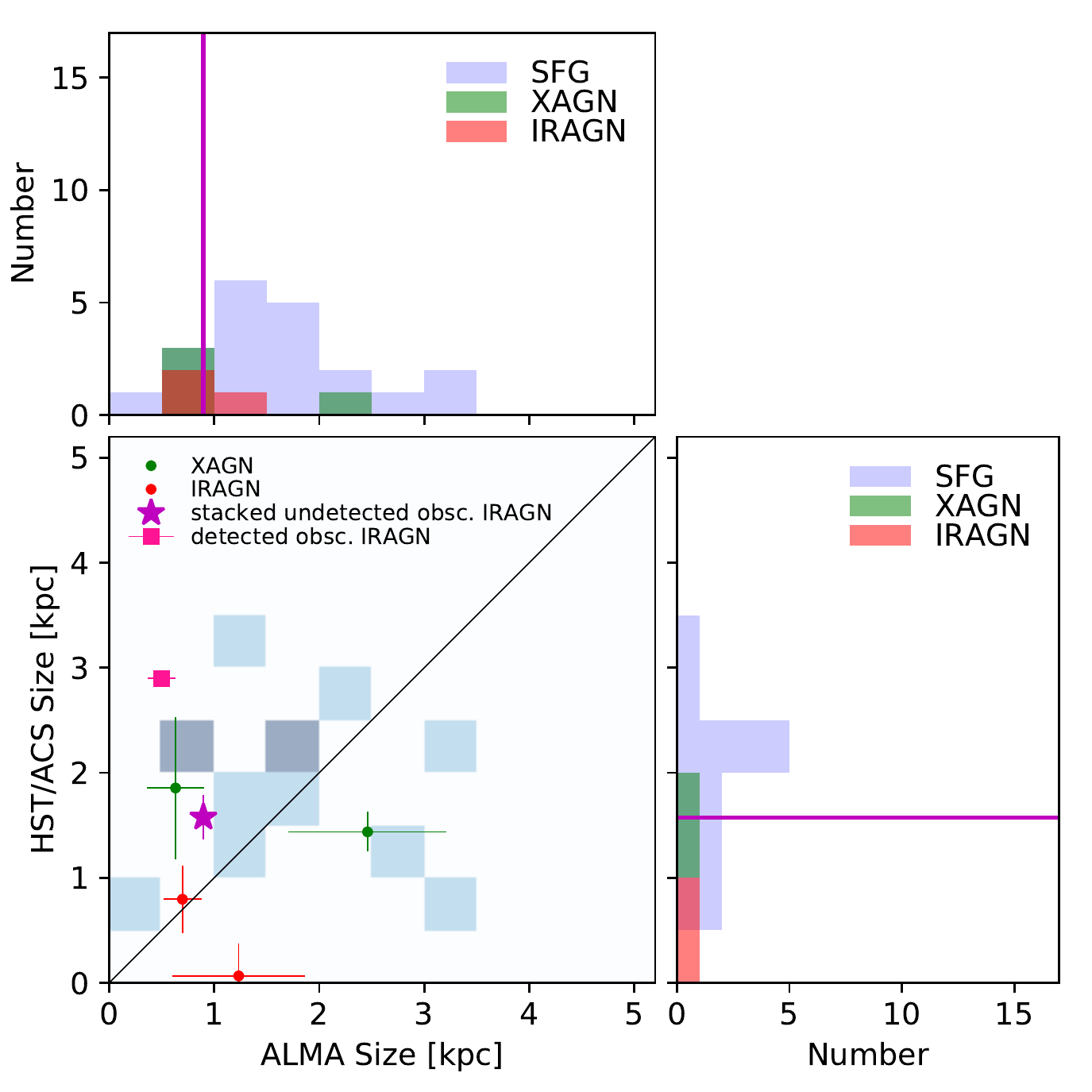} 
\caption{Upper left: histograms of ALMA/1~mm size. Lower right: histogram of HST/$F814W$ size. The magenta lines show the effective radii of the stacked images from the nine undetected obscured AGN source. Here the detected obscured AGN is included in the IRAGN sample. Lower left: HST/$F814W$ size to ALMA/1~mm size plot. The blue color represents the density of SFGs in each size bin. This subplot only includes objects which have measurements of both HST/$F814W$ and ALMA/1~mm  sizes.}
\label{alma_rr}
\end{figure*}

\begin{figure*}
\centering
\includegraphics[width=1.75\columnwidth]{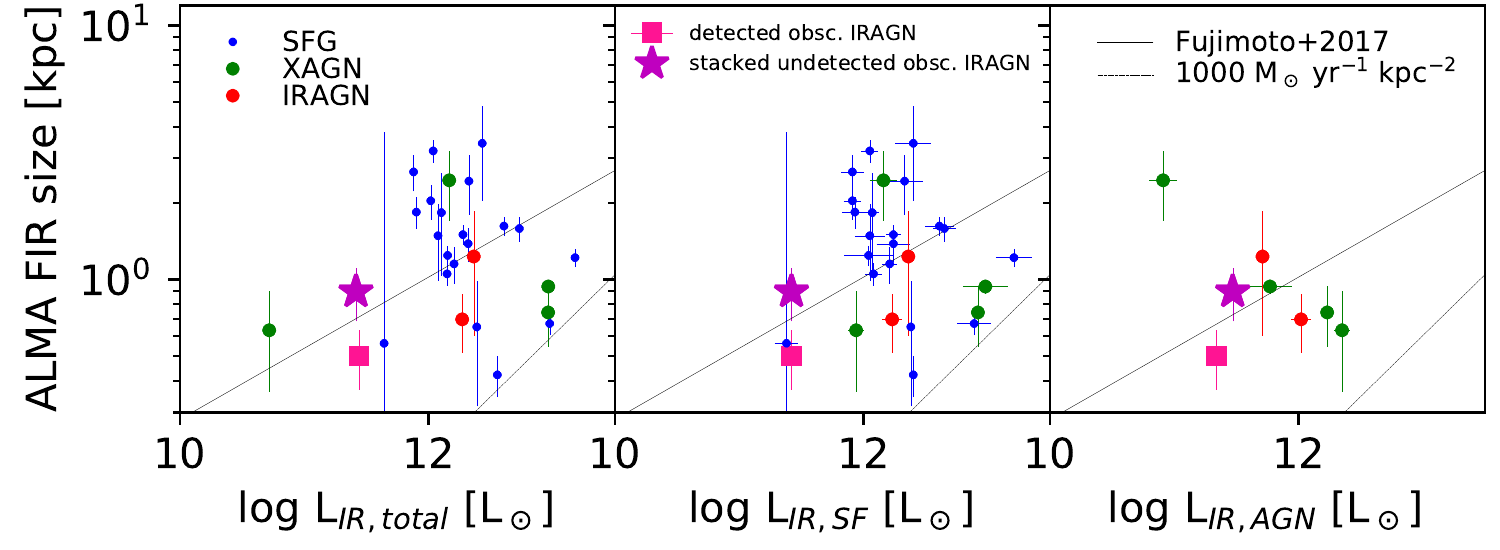} 
\caption{FIR size versus total infrared luminosity, infrared luminosity from star formation and AGN contributions. The black lines are from \citet{2017ApJ...850...83F}. The symbols are the same as described in Figure~\ref{alma_msfr}.}
\label{alma_rl}
\end{figure*}

\begin{figure*}
\centering
\includegraphics[width=1.75\columnwidth]{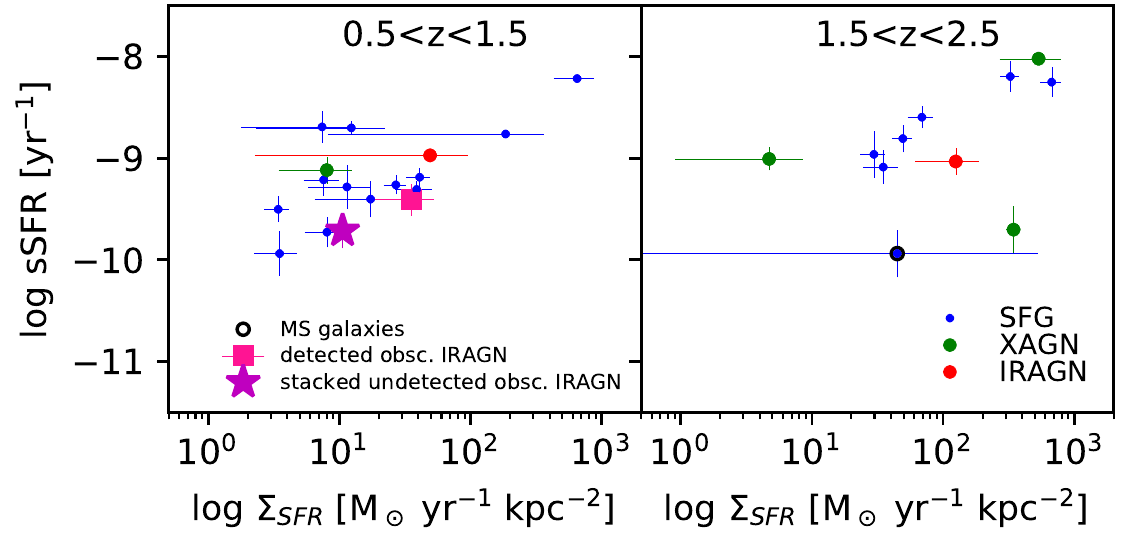} 
\caption{Specific star formation rate to SFR surface density plot in the two redshift bins. Our obscured IRAGN hosts are very compact compared to archival ALMA observations, especially for those on the star-forming sequence. The symbols are the same as described in Figure~\ref{alma_msfr}.}
\label{alma_sigma}
\end{figure*}


\section{Analysis}
\label{sec3}

We separated the sample between non-AGN star-forming galaxies (SFGs), X-ray selected AGNs (XAGNs: $\rm{L}_X$(2-10 keV)$>10^{42}$ ergs/s), and infrared selected AGNs (IRAGNs: using the mid-infrared colors from the IRAC 3.6, 5.8, 4.5, and 8.0 $\mu$m; obscured IRAGNS: $L_{IR,AGN}/L_{X,AGN} > 20$), that are identified by previous work \citep{2016ApJ...819...62C,2016ApJ...817...34M,2017ApJS..233...19C,2017MNRAS.466L.103C}.
Figure~\ref{alma_msfr} shows the SFR versus stellar mass diagrams in the two redshift bins. Most of the sources in the sample from the ALMA archive are massive and have high SFRs, and they are in general on and above the star-forming sequence. 
Our star-forming sequences are estimated by the SED fitting results for ${\rm NUV}-r$ vs. $r-J$ selected galaxies. We also show the main sequence relations from \citet{2014ApJS..214...15S} and \citet{2014ApJ...795..104W}. Our SFR estimations are generally lower than those in the literature by $\sim$ 0.1-0.3 dex at z$\sim$2. The main reason might be selection of the star-forming sample. With the same sample selection, our main sequences are very close ($<$0.1 dex) to the main sequences calculated by the stellar masses and SFRs in the COSMOS2015 catalogs. To avoid bias from different selection criteria, we use the ${\rm NUV}-r$ vs. $r-J$ selection, but show the literature in  Figure\,~\ref{alma_msfr} for comparison. The SFR differences between the various works are more significant at the high stellar mass end, so it is important to compare AGN hosts with non-AGN galaxies using SFRs derived with the same method.
Comparing to the archival data at $z\sim1$, our AGN sample have low SFRs because they were selected to be within +/-0.3 dex of the main sequence. This may be the reason why our sample was mostly undetected with ALMA. 
The only one IRAGN at $z\sim1$ has lower stellar mass ($\sim$0.14 dex) and higher SFR ($\sim$0.20 dex) than non-AGN galaxies.
There is only one IRAGNs at $z\sim2$, with lower stellar masses ($\sim$0.04dex) and lower SFRs ($\sim$0.35dex) than the median of the non-AGN galaxies.

\begin{table}
\begin{tabular}{ccc}
\hline
\hline
type & $z\sim$1 & $z\sim$2\\
\hline
SFG & 1.74$\pm$0.50 (1.62$\pm$0.50) & 1.25$\pm$0.73 (1.30$\pm$0.39)\\
XAGN & 2.46 (one object) & 0.69$\pm$0.25 (0.70$\pm$0.24)\\
IRAGN & 0.77$\pm$0.50 (0.76$\pm$0.48) & 0.70 (one object)\\
\hline
\end{tabular}
\caption{The median (mean) FIR sizes for $\log \rm{M}_* \sim 11$ sources in kpc.}
\label{tab_1}
\end{table}

Figure~\ref{alma_rzm} shows the FIR size as a function of redshift (left), stellar mass (middle), and the distance to the star-forming sequence (right) as shown in Figure~\ref{alma_msfr}. The sizes of our detected (0.43$\pm$0.28 kpc) and stacked (0.90$\pm$0.21 kpc) obscured AGN hosts are very compact compared to those from the archive data (mean=1.76$\pm$0.25 kpc and median=1.67$\pm$0.30 kpc at $z\sim$1 and $\log \rm{M}_* \sim 11$). The uncertainties here and below are estimated from a bootstrapping analysis.
As shown in Table~\ref{tab_1}, for $\log \rm{M}_* \sim 11$ sources at $z\sim$1,  the median (mean) FIR sizes of the SFGs, XAGNs, IRAGNs, are 1.74$\pm$0.50 (1.62$\pm$0.50) 2.46 (one object\footnote{For only one object in the stellar mass and redshift bin, the uncertainties are not available.}, 0.77$\pm$0.50 (0.76$\pm$0.48), respectively. 
As shown in Table~\ref{tab_1}, for $\log \rm{M}_* \sim 11$ sources at $z\sim$2,  the median (mean) FIR sizes of the SFGs, XAGNs, IRAGNs, are 1.25$\pm$0.73 (1.30$\pm$0.39), 0.69$\pm$0.25 (0.70$\pm$0.24), 0.70 (one object), respectively.
We find systematically smaller galaxy sizes at $z\sim$2 relative to the corresponding samples at $z\sim$1. The measurements of HST/$F814W$ images are rest-frame optical sizes for $z\sim$1 sources but rest-frame UV sizes for $z\sim$2 sources \citep[e.g., ][]{2014ApJ...788...28V}, and color gradient which shows redder centers should be also taken into account \citep[e.g., ][]{2012ApJ...753..114W}.
Most importantly, the number of objects in each sample is limited, especially for AGN sample.
Our IRAGN sample were selected to be within +/-0.3 dex of the main sequence, so we also consider $\log \rm{M}_* \sim 11$ SFGs on the main sequence. For $\log \rm{M}_* \sim 11$ SFGs on the main sequence (the distance to the star-forming sequence $<$0.5 dex; a less strict range because of the limited sample size) at $z\sim$1, the median (mean) sizes are 1.97$\pm$0.44 kpc (1.83$\pm$0.49 kpc). For $\log \rm{M}_* \sim 11$ SFGs on the main sequence (the distance to the star-forming sequence $<$0.5 dex) at $z\sim$2, the median (mean) sizes are 1.26$\pm$1.43 kpc (1.27$\pm$1.70 kpc). As a result, we find that IRAGN hosts are in general smaller than SFGs on the main sequence at a given stellar mass and redshift bin.
However, it is difficult to find a reasonable parent sample for our sources on the main sequence because all star-forming galaxies are above the main sequence and mainly in the starburst regime. Therefore, future ALMA observations with better matched control sample are necessary to expand the parameter spaces.

Figure~\ref{alma_rr} shows the histograms of HST/$F814W$, and ALMA/1~mm  sizes. 
As shown in Table~\ref{tab_2}, the mean (median) optical sizes of the SFGs, XAGNs, IRAGNs, are 1.95$\pm$0.74 (2.16$\pm$0.74), 1.58$\pm$0.50 (1.56$\pm$0.50), 0.42$\pm$0.42 (0.41$\pm$0.42), respectively. 
And the mean (median) FIR sizes of the SFGs, XAGNs, IRAGNs, are 1.58$\pm$0.55 (1.48$\pm$0.39), 1.07$\pm$0.52 (0.89$\pm$0.56), 0.75$\pm$0.39 (0.68$\pm$0.45), respectively. 
Considering the uncertainties of the sizes and the different measurement techniques as described in Section~\ref{sec2}, these numbers imply that the rest-frame FIR sizes of all galaxies seem to be similar or more compact than their rest-frame optical sizes.
In particular, the FIR size(0.90$\pm$0.21 kpc)  of our stacked sources is smaller than the optical one (1.58$\pm$0.01).
Though there are few IRAGNs, most of them show both very compact optical and FIR sizes.
The rest-frame optical and FIR sizes of XAGNs are also relatively small or similar as shown in Figure~\ref{alma_rr}.

There are only two objects in the CANDELS field as shown in Table~\ref{tab_0}, and one XAGN is at z$\sim$2. 
The HST/$F160W$ measurements are better tracers for optical sizes than the HST/$F814W$ ones at z$\sim$2. The current data shows that the two sizes in HST/$F160W$ are comparable but larger than most sizes in HST/$F814W$. HST/$F160W$ traces rest-frame the optical size and HST/$F814W$ traces the UV size for the same object, so it is difficult to make a direct comparison. We need size measurements using observations at the same (rest-frame) wavelength for larger numbers of objects to draw a statistical conclusion at $z\sim2$.

\begin{table}
\begin{tabular}{ccc}
\hline
\hline
type & HST/$F814W$ size & ALMA/1~mm size\\
\hline
SFG & 1.95$\pm$0.74 (2.16$\pm$0.74) &  1.58$\pm$0.55 (1.48$\pm$0.39)\\
XAGN & 1.58$\pm$0.50 (1.56$\pm$0.50) & 1.07$\pm$0.52 (0.89$\pm$0.56)\\
IRAGN & 0.42$\pm$0.42 (0.41$\pm$0.42) & 0.75$\pm$0.39 (0.68$\pm$0.45)\\
\hline
\end{tabular}
\caption{The median (mean) optical (HST/$F814W$) and FIR (ALMA/1~mm) sizes in kpc.}
\label{tab_2}
\end{table}

Figure~\ref{alma_rl} shows the ALMA/1~mm sizes to the infrared luminosity relation. 
Our galaxies fall more or less on top of the \citet{2017ApJ...850...83F} line, but it is difficult to show a clear relation because of the scatter and sample size. 
Here we decompose the AGN and star formation components according to the SEDs by MAGPHYS+AGN as described in \citet{2017ApJS..233...19C}.
Therefore, we are able to derive the infrared luminosity from the star formation and AGN components, respectively. 
Because FIR luminosity traces dusts, this decomposition can separate dust emission of AGNs from their host galaxies. 
Figure~\ref{alma_rl} does not show an obvious correlation between the FIR size and the IR luminosity from the AGN component.
According to our SED decomposition, the AGN contribution in the FIR range is negligible ($<$1\% of the total flux), so the ALMA/~1mm sizes are dominated by star-forming populations. Therefore, the middle panel in Figure~\ref{alma_rl} is approximate to the FIR size of hosts versus the infrared luminosity from the star-forming part. It is possible that there are correlations between the FIR sizes of the AGNs and the infrared luminosity from the AGN part. However, deeper and higher resolution data are required to resolve the AGN component at 1mm with ALMA.

Figure~\ref{alma_sigma} shows the specific star formation rate (sSFR=SFR/$M_*$) to SFR surface density ($\Sigma_{SFR}=\rm{SFR}/ (\pi R_e^{2})$, where $R_e$ is the rest-frame FIR size) plots. 
Because there are more massive and star-forming sources at high redshifts, it is clear to see that high redshift sources having higher SFR will naturally imply higher SFR surface density. 
It's also mentioned earlier that the sizes were systematically smaller at higher redshift, which would also increase the surface densities even at a fixed SFR or a fixed sSFR.
At $z\sim1$, the SFR surface density of our obscured IRAGN hosts are high compared to archived ALMA observations at a given sSFR. At $z\sim2$, the SFR surface density of both XAGN and IRAGN hosts seem to be similar or slightly higher than that of SFGs, but the uncertainties of the SFR surface density are large. 
In Figure~\ref{alma_sigma}, SFR surface densities of our IRAGN hosts are much higher than SFGs on the main sequence at both $z\sim1$. This suggests that the compactness of IRAGN hosts will be significant if we have a large control sample of SFGs.
Therefore, more data with high resolution are needed to reach a statistically meaningful conclusion.



\section{Discussion}
\label{sec4}

\subsection{Compact Rest-Frame FIR and Optical sizes}
We used high resolution ALMA images to measure the rest-frame FIR sizes, and compared the dust maps with the \emph{HST} optical images for non-AGN star-forming galaxies, X-ray selected AGNs, and IR selected AGNs. 
We found that the rest-frame FIR sizes of IRAGN hosts are as compact ($<$1kpc) as their rest-frame optical sizes, which suggests that they are indeed compact star-forming galaxies which have nuclear activities due to compaction \citep[e.g.,][]{2015MNRAS.450.2327Z,2015MNRAS.452.1502D,2016MNRAS.457.2790T}.  
It is possible that the detected sources may be more concentrated galaxies than undetected ones because we may not detect galaxies that are too extended in the high-resolution ALMA observations given the sensitivity limit. However, we apply the analyses equally to AGN and non-AGN galaxies, so the conclusion would not be changed. 
A possible explanation might be that obscured AGNs are triggered in compact galaxies. Such compact profiles could also be explained if obscured AGNs are caught at a special moment when their host galaxies still harbor intense star formation but simultaneously experience a phase of compaction, as what was recently suggested from simulations of gas-rich galaxies undergoing violent disk instabilities. The possible migration of their giant star-forming clumps toward their central region could lead to enhanced AGN fueling \citep{2012ApJ...757...81B,2016MNRAS.457.2790T,2015MNRAS.452.1502D} as well as to the quick formation of a dense stellar component \citep{2009Natur.457..451D,2010MNRAS.404.2151C}.
According to \citet{2016MNRAS.457.2790T}, post-compaction blue nuggets are expected to be found above the main-sequence ridge. 
Most of the archival sources seem to be biased on the higher side of the main sequence and above, so there is an important selection effect here when comparing the samples. 
Our sources were selected to lie on the MS relation within +/- 0.3 dex (Figure~\ref{alma_msfr}), so by selection, they do not have the same bias toward higher SFR.  That might also explain why they do not lie in the expected blue nugget location.
The lack of comparison sample at the same stellar mass, SFR, and redshift bin might be also the reason why we did not see much higher SFR surface density of our obscured AGNs in Figure~\ref{alma_sigma}. 
Our results may imply that the dusty star formation of AGN hosts happens in compact regions, and forms the stellar component of the host galaxy. However, more FIR data with better matched control sample are needed to confirm the simulation predictions.

\subsection{FIR versus Optical Sizes in AGN Hosts}
It has been found that the rest-frame FIR sizes are in general smaller than the rest-frame optical sizes \citep[e.g.,][]{2016ApJ...827L..32B, 2017ApJ...841L..25T, 2017ApJ...850...83F, 2018ApJ...861....7F, 2018A&A...616A.110E}, which can be explained by compact dusty starbursts.
Though the size estimation can be sensitive to the measurement methods, model profiles, and initial values \citep[e.g.,][]{2018ApJ...861....7F,2018A&A...616A.110E}, the main results of our analysis do not change even if we use different ways to measure the sizes.
We cross-matched our sample with \citet{2018ApJ...861....7F}, who showed consistent size measurement with previous ALMA results \citep{2015ApJ...807..128S,2015ApJ...810..133I,2016ApJ...833..103H,2016ApJ...827L..32B,2016ApJ...833...12R,2017ApJ...841L..25T,2017A&A...597A..41G}. 
We found 20 matches out of our 27 sources. Among them, there is no significant difference (systematic difference $\sim$0.05$\pm$0.10kpc; our value is smaller but close to \citet{2018ApJ...861....7F}) in between our size estimation with \texttt{UVMODELFIT} and their FWHM estimated with the \texttt{UVMULTIFIT} task of CASA.
Moreover, \citet{2018A&A...616A.110E} show that the semi-major axis from uv (\texttt{UVMODELFIT} in CASA) and image (\texttt{GALFIT}) planes agree with only a 5\% systematic difference.
Therefore, our size measurements should be robust against those of other size measurement methods. 
A more sophisticated analysis might be performed with GALFIT \citep[e.g., bulge-disk decomposition for the PHIBSS2 galaxies; ][]{2019A&A...622A.105F,2018ApJ...853..179T}, but more HST/$F160W$ images or higher resolution ALMA images ($\sim0\arcsec.1$) of the complementary sample at $z<1$ are required. 
Our results show that the rest-frame FIR sizes of both AGN and SFG seem to be similar or more compact than their rest-frame optical sizes. 
This may suggest that the star formation mechanisms in the compact regions of AGN hosts are similar to those of other ALMA detected sources. However, more detected AGN hosts with ALMA observations are needed to confirm this result.

\subsection{SFR Surface Densities of Obscured IRAGNs}
Our galaxies fall more or less on top of the \citet{2017ApJ...850...83F} line, but it is difficult to show a clear relation because of the scatter and sample size. 
Our measurements are also above the maximum SFR surface density ($\sim$1000$M_\odot \rm{yr}^{-1} \rm{kpc}^{-2}$ as shown in Figure~\ref{alma_sigma}) which was estimated by \citet{2015ApJ...807..128S}. The decomposed IRAGN luminosity may imply that AGN activity has little influence on the FIR sizes. However, more high-resolution imaging from future ALMA observations are required to show a statistical result.
On the other hand, the median SFR surface density of the obscured AGN hosts is more than two times higher than those of other ALMA-detected sources at $\log \rm{M}_*/M_\odot \sim 11$ and $z\sim1$.
This implies that obscured AGN host galaxies may be in a transition phase of compact star-forming galaxies, and may evolve to compact quiescent galaxies at later time \citep{2015MNRAS.450.2327Z,2015MNRAS.452.1502D}.
The obscured black hole accretion phase may be related to galaxies experiencing a compaction of their gaseous component, which could play a key role in the final quenching mechanisms driving the formation of passive red sources.
Nevertheless, there are still very few complementary samples with matched stellar mass, SFR, and redshift. Most of the archival ALMA observations focus on massive galaxies above the star-forming sequence. Therefore, future high-resolution ALMA observations would be crucial to investigating further about these compact AGN host galaxies. 



\section{Summary}
\label{sec5}

In this paper, we have investigated rest-frame FIR and optical sizes of non-AGN star-forming galaxies, infrared selected AGNs, and X-ray selected AGNs at $z<2.5$ by using high-resolution ALMA ($0\arcsec.10$-$0\arcsec.85$) and HST imaging ($\sim0\arcsec.1$) in the COSMOS field. Our main findings are as follows.

\begin{enumerate}

\item The sizes measured from both ALMA/1~mm  and HST/$F814W$ images show that obscured AGN host galaxies at z$\sim$1 are more compact than star-forming galaxies at a given redshift and stellar mass.

\item Most detected rest-frame FIR sizes of the sources are similar or more compact than their rest-frame optical sizes, which is consistent with recent results for ALMA detected sources. This can be explained by the fact that the dusty starbursts take place in the compact regions, and suggests that the star formation mechanisms in the compact regions of AGN hosts are similar to those taking place in star-forming galaxies with no evidence of an AGN at their center.

\item The decomposed IRAGN luminosity may imply that AGN activity has little influence on the FIR sizes. However, more high-resolution imaging from future ALMA observations are needed to confirm this result.

\item The median SFR surface density of the obscured AGN hosts is higher than those of other ALMA-detected sources at $\log \rm{M}_*/M_\odot \sim 11$ and $z\sim1$. This implies that obscured AGN host galaxies may be in a transition phase of compact star-forming galaxies, and may evolve into compact quiescent galaxies at later time. 

\item Future high-resolution ALMA observations would be crucial to investigating further about these compact AGN host galaxies.

\end{enumerate}


\acknowledgments
We thank the referee as well as J. Freundlich for helpful comments and discussions.
YYC acknowledge financial support from the Ministry of Science and Technology of Taiwan grant (105-2112-M-001-029-MY3 and 108-2112-M-001-014-), and the Agence Nationale de la Recherche (contract \#ANR-12-JS05-0008-01). 
EdC gratefully acknowledges the Australian Research Council as the recipient of a Future Fellowship (project FT150100079).
This paper makes use of the following ALMA data: ADS/JAO.ALMA \#2013.1.00034.S, \#2015.1.00026.S, \#2016.1.00478.S, \#2016.1.00735.S, \#2016.1.00778.S, \#2016.1.01355.S, \#2016.1.01426.S, and \#2016.1.01604.S. ALMA is a partnership of ESO (representing its member states), NSF (USA) and NINS (Japan), together with NRC (Canada), MOST and ASIAA (Taiwan), and KASI (Republic of Korea), in cooperation with the Republic of Chile. The Joint ALMA Observatory is operated by ESO, AUI/NRAO and NAOJ.
We gratefully acknowledge the contributions of the entire COSMOS collaboration.





\clearpage
\begin{sidewaystable}
\vspace*{5cm}
\centering
\tiny
\setlength{\tabcolsep}{0.5pt}
\renewcommand{\arraystretch}{1.0} 
\begin{tabular}{cccccccccccccccccccccc}
\hline
\hline
Project Code & Source Name & Band & Beam Size & $\rm SN_{\rm peak}$ & $\rm S_{\rm peak}$  & $\rm S_{\rm int}$ & $\rm ID_{\rm COSMOS}$ & ra & dec & $z$ & $\rm M_*$ & SFR & R$_{\rm ACS(g)}$ & R$_{\rm WFC3(g)}$ & R$_{\rm ALMA(g)}$ & R$_{\rm ACS(c)}$ & R$_{\rm ALMA(c)}$ & R$_{\rm ALMA(d)}$ & R$_{\rm ALMA(UV)}$ & AGN \\
 & & & [$\arcsec \times \arcsec$] & & [mJy/beam] & [mJy] & [deg] & [deg] & & [$M_\odot$] & [$M_\odot\rm{yr}^{-1}$] & [kpc] & [kpc] & [kpc] & [kpc] & [kpc] & [kpc] & [kpc] & \\
 (1) & (2) & (3) & (4) & (5) & (6) & (7) & (8) & (9) & (10) & (11) & (12) & (13) & (14) & (15) & (16) & (17) & (18) & (19) & (20) & (21) \\
\hline
2013.1.00034.S & lowz\_cell8\_6496 & 7 & 0.33$\times$0.24 & 3.54 & 0.39$\pm$0.11 & 2.17$\pm$0.72 & 239908 & 149.97412 & 1.64352 & 1.04 & 11.2$\pm$0.1 & 2.2$\pm$0.0 & 1.4$\pm$0.2 & - & - & 0.6$\pm$0.0 & 2.7$\pm$0.5 & 2.4$\pm$0.8 & 2.5$\pm$0.8 & XAGN \\
2016.1.01604.S & 317108 & 7 & 0.39$\times$0.39 & 14.61 & 0.71$\pm$0.05 & 1.64$\pm$0.15 & 317108 & 150.15256 & 1.76921 & 1.56 & 11.2$\pm$0.1 & 2.3$\pm$0.1 & - & - & 1.9$\pm$0.1 & 3.6$\pm$0.1 & 2.5$\pm$0.1 & 1.8$\pm$0.2 & 1.4$\pm$0.2 & - \\
2016.1.01604.S & 360903 & 7 & 0.34$\times$0.29 & 5.67 & 0.30$\pm$0.05 & 0.85$\pm$0.20 & 360903 & 149.66491 & 1.83649 & 1.35 & 11.5$\pm$0.1 & 1.9$\pm$0.1 & 1.2$\pm$0.6 & - & - & 3.6$\pm$0.0 & 2.2$\pm$0.3 & 1.7$\pm$0.4 & 2.7$\pm$0.4 & - \\
2016.1.01604.S & 405986 & 7 & 0.33$\times$0.29 & 8.04 & 0.36$\pm$0.04 & 2.03$\pm$0.29 & 405986 & 149.79795 & 1.90880 & 1.35 & 11.3$\pm$0.1 & 2.0$\pm$0.1 & 0.6$\pm$0.5 & - & 3.0$\pm$0.1 & 2.4$\pm$0.0 & 3.2$\pm$0.3 & 2.8$\pm$0.3 & 3.2$\pm$0.3 & - \\
2016.1.01604.S & 495291 & 7 & 0.39$\times$0.39 & 26.43 & 1.07$\pm$0.04 & 1.75$\pm$0.10 & 495291 & 150.62793 & 2.05289 & 1.30 & 11.1$\pm$0.1 & 2.2$\pm$0.1 & 3.4$\pm$0.7 & - & - & 3.1$\pm$0.1 & 2.1$\pm$0.1 & 1.3$\pm$0.1 & 1.1$\pm$0.1 & - \\
2016.1.01604.S & 500431 & 7 & 0.39$\times$0.39 & 13.05 & 0.53$\pm$0.04 & 1.43$\pm$0.15 & 500431 & 150.13314 & 2.06073 & 1.39 & 11.0$\pm$0.1 & 1.9$\pm$0.1 & - & - & 2.5$\pm$0.9 & 3.5$\pm$0.1 & 2.7$\pm$0.2 & 1.8$\pm$0.3 & 1.8$\pm$0.3 & - \\
2013.1.00034.S & lowz\_cell11\_120263 & 7 & 0.40$\times$0.24 & 9.55 & 1.73$\pm$0.18 & 2.92$\pm$0.45 & 504172 & 149.83252 & 2.06602 & 1.14 & 11.4$\pm$0.0 & 2.4$\pm$0.0 & 2.2$\pm$0.6 & - & - & 2.8$\pm$0.0 & 1.6$\pm$0.1 & 0.8$\pm$0.3 & 0.7$\pm$0.3 & - \\
2016.1.00735.S & CXID-36 & 7 & 0.11$\times$0.10 & 15.19 & 0.66$\pm$0.04 & 2.69$\pm$0.22 & 553928 & 150.15841 & 2.13958 & 1.83 & 11.8$\pm$0.7 & 3.0$\pm$0.0 & 0.0$\pm$8.3 & - & - & 0.4$\pm$0.0 & 0.9$\pm$0.0 & 0.8$\pm$0.1 & 0.9$\pm$0.1 & XAGN \\
2016.1.01426.S & PACS\_819 & 6 & 0.13$\times$0.09 & 13.47 & 0.14$\pm$0.01 & 0.56$\pm$0.05 & 628146 & 149.98145 & 2.25320 & 1.40 & 10.8$\pm$0.0 & 2.6$\pm$0.0 & 1.0$\pm$0.4 & - & - & 7.0$\pm$0.1 & 0.9$\pm$0.0 & 0.8$\pm$0.1 & 0.4$\pm$0.1 & - \\
2016.1.01604.S & 659927 & 7 & 0.34$\times$0.29 & 8.05 & 0.41$\pm$0.05 & 1.17$\pm$0.19 & 659927 & 150.34101 & 2.30099 & 1.44 & 11.4$\pm$0.1 & 2.0$\pm$0.1 & 2.6$\pm$0.6 & - & - & 3.8$\pm$0.0 & 2.3$\pm$0.2 & 1.6$\pm$0.3 & 2.0$\pm$0.3 & - \\
2016.1.01604.S & 766157 & 7 & 0.33$\times$0.29 & 13.39 & 0.84$\pm$0.06 & 1.51$\pm$0.16 & 766157 & 150.00352 & 2.46146 & 1.40 & 11.3$\pm$0.1 & 2.2$\pm$0.0 & 1.1$\pm$101.3 & - & - & 2.5$\pm$0.1 & 1.8$\pm$0.1 & 1.2$\pm$0.2 & 1.2$\pm$0.2 & - \\
2013.1.00034.S & midz\_cell8\_254150 & 7 & 0.33$\times$0.30 & 15.95 & 2.37$\pm$0.15 & 3.09$\pm$0.31 & 794848 & 150.09342 & 2.50734 & 2.20 & 10.8$\pm$0.1 & 0.8$\pm$0.2 & 1.9$\pm$0.7 & 3.1$\pm$0.1 & - & 2.0$\pm$0.1 & 1.5$\pm$0.1 & 0.7$\pm$0.3 & 0.6$\pm$0.3 & XAGN \\
2016.1.00778.S & J100054.83+023126.2 & 6 & 0.45$\times$0.32 & 9.68 & 0.33$\pm$0.03 & 0.48$\pm$0.08 & 806579 & 150.23118 & 2.52466 & 1.95 & 10.9$\pm$0.1 & 1.6$\pm$0.2 & 22.3$\pm$22.4 & - & - & - & 1.9$\pm$0.1 & 0.6$\pm$0.7 & 0.6$\pm$3.3 & - \\
2013.1.00034.S & lowz\_cell3\_258881 & 7 & 0.52$\times$0.36 & 7.44 & 0.97$\pm$0.13 & 1.64$\pm$0.33 & 811432 & 150.04247 & 2.52656 & 1.23 & 11.2$\pm$0.1 & 2.1$\pm$0.0 & 1.8$\pm$1.1 & - & 2.3$\pm$1.0 & 4.5$\pm$0.1 & 2.4$\pm$0.2 & 1.2$\pm$0.7 & 1.5$\pm$0.5 & - \\
2013.1.00034.S & midz\_cell11\_262500 & 7 & 0.37$\times$0.34 & 26.26 & 3.72$\pm$0.14 & 6.70$\pm$0.37 & 815012 & 150.60329 & 2.53654 & 2.19 & 11.3$\pm$0.1 & 3.2$\pm$0.0 & 1.1$\pm$2.0 & - & - & 0.8$\pm$0.0 & 2.0$\pm$0.1 & 1.3$\pm$0.1 & 1.2$\pm$0.1 & - \\
2016.1.01604.S & 816106 & 7 & 0.39$\times$0.39 & 19.19 & 0.86$\pm$0.05 & 1.81$\pm$0.13 & 816106 & 150.08329 & 2.53609 & 1.36 & 11.3$\pm$0.1 & 2.3$\pm$0.1 & 2.3$\pm$0.7 & 3.5$\pm$0.0 & - & 3.2$\pm$0.0 & 2.4$\pm$0.1 & 1.7$\pm$0.1 & 1.5$\pm$0.1 & - \\
2013.1.00034.S & lowz\_cell11\_268592 & 7 & 0.53$\times$0.36 & 3.45 & 0.36$\pm$0.10 & 2.24$\pm$0.75 & 831023 & 150.41785 & 2.55852 & 1.21 & 11.0$\pm$0.1 & 2.4$\pm$0.0 & 2.5$\pm$0.3 & - & - & 2.7$\pm$0.0 & 4.5$\pm$0.9 & 4.1$\pm$1.2 & 3.4$\pm$1.4 & - \\
2015.1.00026.S & SHIZELS-14 & 6 & 0.25$\times$0.21 & 12.96 & 0.24$\pm$0.02 & 2.33$\pm$0.20 & 831167 & 150.21496 & 2.55952 & 2.08 & 11.2$\pm$0.1 & 2.7$\pm$0.0 & 2.0$\pm$0.8 & - & - & 2.3$\pm$0.0 & 3.0$\pm$0.2 & 2.7$\pm$0.2 & 1.6$\pm$0.2 & - \\
2016.1.00478.S & AzTECC2a & 7 & 0.20$\times$0.17 & 7.72 & 1.26$\pm$0.16 & 1.78$\pm$0.36 & 842595 & 149.99797 & 2.57823 & 2.42 & 10.8$\pm$0.0 & 3.0$\pm$0.0 & 0.7$\pm$175.9 & - & - & 3.6$\pm$0.1 & 0.9$\pm$0.1 & 0.5$\pm$0.2 & 0.7$\pm$0.2 & XAGN \\
2013.1.00034.S & lowz\_cell11\_286398 & 7 & 0.54$\times$0.37 & 6.84 & 0.78$\pm$0.11 & 4.03$\pm$0.70 & 872762 & 150.12960 & 2.62144 & 1.41 & 11.4$\pm$0.1 & 2.3$\pm$0.1 & - & - & - & 8.2$\pm$0.2 & 4.3$\pm$0.4 & 3.6$\pm$0.6 & 2.4$\pm$0.6 & - \\
2016.1.01604.S & 888769 & 7 & 0.39$\times$0.39 & 23.00 & 1.02$\pm$0.04 & 1.62$\pm$0.11 & 888769 & 150.10745 & 2.64591 & 1.54 & 10.9$\pm$0.2 & 2.2$\pm$0.1 & - & - & 2.0$\pm$1.1 & 1.5$\pm$0.0 & 2.1$\pm$0.1 & 1.3$\pm$0.1 & 1.2$\pm$0.1 & - \\
2016.1.00478.S & AzTECC22a & 7 & 0.19$\times$0.18 & 19.99 & 3.56$\pm$0.18 & 6.41$\pm$0.47 & 902320 & 150.03729 & 2.66960 & 1.85 & 11.0$\pm$0.1 & 3.0$\pm$0.0 & 2.3$\pm$0.6 & - & - & 1.9$\pm$0.0 & 1.0$\pm$0.0 & 0.7$\pm$0.1 & 0.7$\pm$0.1 & - \\
2016.1.01604.S & 917430 & 7 & 0.34$\times$0.29 & 13.33 & 0.70$\pm$0.05 & 0.86$\pm$0.11 & 917430 & 149.91774 & 2.69211 & 1.60 & 11.1$\pm$0.1 & 2.3$\pm$0.1 & 0.8$\pm$0.3 & - & - & 0.6$\pm$0.0 & 1.5$\pm$0.1 & 0.6$\pm$0.2 & 0.7$\pm$0.2 & IRAGN \\
2013.1.00034.S & lowz\_cell10\_324567 & 7 & 0.56$\times$0.38 & 5.53 & 0.73$\pm$0.13 & 1.45$\pm$0.38 & 960580 & 149.91909 & 2.76053 & 1.13 & 10.7$\pm$0.1 & 2.1$\pm$0.0 & 2.3$\pm$0.3 & - & - & 3.0$\pm$0.0 & 2.7$\pm$0.4 & 1.9$\pm$0.7 & 1.8$\pm$0.8 & - \\
2013.1.00034.S & lowz\_cell10\_328266 & 7 & 0.57$\times$0.38 & 5.81 & 0.80$\pm$0.14 & 1.61$\pm$0.40 & 969105 & 150.53066 & 2.77548 & 1.35 & 11.2$\pm$0.0 & 2.4$\pm$0.0 & 0.1$\pm$0.3 & - & - & 0.6$\pm$0.0 & 2.8$\pm$0.3 & 2.0$\pm$0.6 & 1.2$\pm$0.6 & IRAGN \\
2016.1.01355.S & 972308 & 7 & 0.12$\times$0.10 & 5.79 & 0.52$\pm$0.09 & 1.37$\pm$0.31 & 972308 & 150.40459 & 2.78053 & 0.60 & 10.6$\pm$0.1 & 1.4$\pm$0.0 & 0.0$\pm$0.0 & - & - & 0.4$\pm$0.0 & 0.6$\pm$0.1 & 0.5$\pm$0.1 & 0.5$\pm$0.1 & IRAGN \\
2013.1.00034.S & midz\_cell6\_332275 & 7 & 0.37$\times$0.33 & 19.46 & 3.00$\pm$0.15 & 7.19$\pm$0.50 & 980250 & 150.01611 & 2.79235 & 1.76 & 11.3$\pm$0.1 & 2.6$\pm$0.0 & - & - & - & 0.7$\pm$0.0 & 2.3$\pm$0.1 & 1.7$\pm$0.1 & 1.6$\pm$0.1 & - \\
\hline
2016.1.01355.S & 263569 & 7 & 0.12$\times$0.10 & - & - & - & 263569 & 149.95048 & 1.68282 & 0.78 & 10.8$\pm$0.1 & 1.3$\pm$0.1 & 2.4$\pm$0.4 & - & - & - & - & - & -& IRAGN \\
2016.1.01355.S & 334665 & 7 & 0.12$\times$0.10 & - & - & - & 334665 & 150.43685 & 1.79439 & 0.56 & 10.9$\pm$0.1 & 1.2$\pm$0.1 & 1.8$\pm$0.1 & - & - & - & - & - & -& IRAGN \\
2016.1.01355.S & 335028 & 7 & 0.12$\times$0.10 & - & - & - & 335028 & 149.50325 & 1.79492 & 0.65 & 11.3$\pm$0.1 & 1.8$\pm$0.1 & 1.5$\pm$0.1 & - & - & - & - & - & -& IRAGN \\
2016.1.01355.S & 410490 & 7 & 0.12$\times$0.10 & - & - & - & 410490 & 150.02499 & 1.91477 & 0.97 & 11.2$\pm$0.1 & 1.9$\pm$0.0 & - & - & - & - & - & - & -& IRAGN \\
2016.1.01355.S & 414248 & 7 & 0.12$\times$0.10 & - & - & - & 414248 & 149.84769 & 1.92058 & 0.71 & 10.7$\pm$0.1 & 1.2$\pm$0.2 & 0.0$\pm$0.0 & - & - & - & - & - & -& IRAGN \\
2016.1.01355.S & 417645 & 7 & 0.12$\times$0.10 & - & - & - & 417645 & 150.02586 & 1.92639 & 0.69 & 10.7$\pm$0.1 & 1.4$\pm$0.1 & 0.7$\pm$0.1 & - & - & - & - & - & -& IRAGN \\
2016.1.01355.S & 456808 & 7 & 0.12$\times$0.10 & - & - & - & 456808 & 149.57039 & 1.99058 & 1.00 & 11.2$\pm$0.1 & 1.1$\pm$0.2 & 1.1$\pm$0.2 & - & - & - & - & - & -& IRAGN \\
2016.1.01355.S & 711692 & 7 & 0.12$\times$0.10 & - & - & - & 711692 & 149.99343 & 2.37723 & 0.93 & 10.7$\pm$0.1 & 1.6$\pm$0.1 & 0.9$\pm$0.3 & - & - & - & - & - & -& IRAGN \\
2016.1.01355.S & 755017 & 7 & 0.12$\times$0.10 & - & - & - & 755017 & 150.76449 & 2.44222 & 0.70 & 10.9$\pm$0.1 & 1.8$\pm$0.1 & 3.4$\pm$0.2 & - & - & - & - & - & -& IRAGN \\
\hline
\end{tabular} \\
{\tiny
\begin{flushleft}
Note:
(1) ALMA project code.
(2) ALMA source name.
(3) ALMA bands.
(4) ALMA beam size. 
(5) ALMA S/N. 
(6) ALMA peak flux from \texttt{IMFIT}.
(7) Integrated total flux from \texttt{IMFIT}.
(8) COSMOS ID.
(9) COSMOS ra.
(10) COSMOS dec.
(11) Best redshift in COSMOS.
(12) Stellar mass estimation from MAGPHYS. The uncertainties of stellar masses are from the probability density functions, which depend on the model templates. The real uncertainty can be larger than our estimation. 
(13) SFR estimation from MAGPHYS. 
(14) Sizes measured from HST/$F814W$ images with \texttt{GALFIT}.
(15) Sizes measured from HST/$F160W$ images with \texttt{GALFIT}
(16) Sizes measured from HST/$F160W$ images with \texttt{GALFIT}
(17) Sizes measured from HST/$F814W$ images with \texttt{IMFIT} (convolved).
(18) Sizes measured from ALMA/~1mm images with \texttt{IMFIT} (convolved).
(19) Sizes measured from ALMA/~1mm images with \texttt{IMFIT} (deconvolved).
(20) Sizes measured from ALMA/~1mm images with \texttt{UVMODELFIT}.
(21) IRAGN, XAGN, or non-AGN galaxies.
\end{flushleft}
}
\caption{Properties of 27 detected (upper) and 9 undetected (lower) ALMA band 6/7 high resolution ($0\arcsec.1$-$0\arcsec.4$) sources. }
\label{tab_0}
\end{sidewaystable}
\clearpage

\bibliographystyle{aasjournal}

\begin{thebibliography}{}

\bibitem[{{Barro} {et~al.}(2016){Barro}, {Kriek}, {P{\'e}rez-Gonz{\'a}lez},
  {Trump}, {Koo}, {Faber}, {Dekel}, {Primack}, {Guo}, {Kocevski},
  {Mu{\~n}oz-Mateos}, {Rujopakarn}, \& {Seth}}]{2016ApJ...827L..32B}
{Barro}, G., {Kriek}, M., {P{\'e}rez-Gonz{\'a}lez}, P.~G., {et~al.} 2016,
  \apjl, 827, L32

\bibitem[{{Bournaud} {et~al.}(2012){Bournaud}, {Juneau}, {Le Floc'h},
  {Mullaney}, {Daddi}, {Dekel}, {Duc}, {Elbaz}, {Salmi}, \&
  {Dickinson}}]{2012ApJ...757...81B}
{Bournaud}, F., {Juneau}, S., {Le Floc'h}, E., {et~al.} 2012, \apj, 757, 81

\bibitem[{{Brinchmann} {et~al.}(2004){Brinchmann}, {Charlot}, {White},
  {Tremonti}, {Kauffmann}, {Heckman}, \& {Brinkmann}}]{2004MNRAS.351.1151B}
{Brinchmann}, J., {Charlot}, S., {White}, S.~D.~M., {et~al.} 2004, \mnras, 351,
  1151

\bibitem[{{Ceverino} {et~al.}(2010){Ceverino}, {Dekel}, \&
  {Bournaud}}]{2010MNRAS.404.2151C}
{Ceverino}, D., {Dekel}, A., \& {Bournaud}, F. 2010, \mnras, 404, 2151

\bibitem[{{Chabrier}(2003)}]{2003PASP..115..763C}
{Chabrier}, G. 2003, \pasp, 115, 763

\bibitem[{{Chang} {et~al.}(2015){Chang}, {van der Wel}, {da Cunha}, \&
  {Rix}}]{2015ApJS..219....8C}
{Chang}, Y.-Y., {van der Wel}, A., {da Cunha}, E., \& {Rix}, H.-W. 2015, \apjs,
  219, 8

\bibitem[{{Chang} {et~al.}(2017{\natexlab{a}}){Chang}, {Le Floc'h}, {Juneau},
  {da Cunha}, {Salvato}, {Civano}, {Marchesi}, {Ilbert}, {Toba}, {Lim}, {Tang},
  {Wang}, {Ferraro}, {Urry}, {Griffiths}, \&
  {Kartaltepe}}]{2017ApJS..233...19C}
{Chang}, Y.-Y., {Le Floc'h}, E., {Juneau}, S., {et~al.} 2017{\natexlab{a}},
  \apjs, 233, 19

\bibitem[{{Chang} {et~al.}(2017{\natexlab{b}}){Chang}, {Le Floc'h}, {Juneau},
  {da Cunha}, {Salvato}, {Civano}, {Marchesi}, {Gabor}, {Ilbert}, {Laigle},
  {McCracken}, {Hsieh}, \& {Capak}}]{2017MNRAS.466L.103C}
---. 2017{\natexlab{b}}, \mnras, 466, L103

\bibitem[{{Chen} {et~al.}(2015){Chen}, {Smail}, {Swinbank}, {Simpson}, {Ma},
  {Alexander}, {Biggs}, {Brandt}, {Chapman}, {Coppin}, {Danielson},
  {Dannerbauer}, {Edge}, {Greve}, {Ivison}, {Karim}, {Menten}, {Schinnerer},
  {Walter}, {Wardlow}, {Wei{\ss}}, \& {van der Werf}}]{2015ApJ...799..194C}
{Chen}, C.-C., {Smail}, I., {Swinbank}, A.~M., {et~al.} 2015, \apj, 799, 194

\bibitem[{{Cisternas} {et~al.}(2011){Cisternas}, {Jahnke}, {Inskip},
  {Kartaltepe}, {Koekemoer}, {Lisker}, {Robaina}, {Scodeggio}, {Sheth},
  {Trump}, {Andrae}, {Miyaji}, {Lusso}, {Brusa}, {Capak}, {Cappelluti},
  {Civano}, {Ilbert}, {Impey}, {Leauthaud}, {Lilly}, {Salvato}, {Scoville}, \&
  {Taniguchi}}]{2011ApJ...726...57C}
{Cisternas}, M., {Jahnke}, K., {Inskip}, K.~J., {et~al.} 2011, \apj, 726, 57

\bibitem[{{Civano} {et~al.}(2016){Civano}, {Marchesi}, {Comastri}, {Urry},
  {Elvis}, {Cappelluti}, {Puccetti}, {Brusa}, {Zamorani}, {Hasinger},
  {Aldcroft}, {Alexander}, {Allevato}, {Brunner}, {Capak}, {Finoguenov},
  {Fiore}, {Fruscione}, {Gilli}, {Glotfelty}, {Griffiths}, {Hao}, {Harrison},
  {Jahnke}, {Kartaltepe}, {Karim}, {LaMassa}, {Lanzuisi}, {Miyaji}, {Ranalli},
  {Salvato}, {Sargent}, {Scoville}, {Schawinski}, {Schinnerer}, {Silverman},
  {Smolcic}, {Stern}, {Toft}, {Trakhenbrot}, {Treister}, \&
  {Vignali}}]{2016ApJ...819...62C}
{Civano}, F., {Marchesi}, S., {Comastri}, A., {et~al.} 2016, \apj, 819, 62

\bibitem[{{da Cunha} {et~al.}(2008){da Cunha}, {Charlot}, \&
  {Elbaz}}]{2008MNRAS.388.1595D}
{da Cunha}, E., {Charlot}, S., \& {Elbaz}, D. 2008, \mnras, 388, 1595

\bibitem[{{da Cunha} {et~al.}(2015){da Cunha}, {Walter}, {Smail}, {Swinbank},
  {Simpson}, {Decarli}, {Hodge}, {Weiss}, {van der Werf}, {Bertoldi},
  {Chapman}, {Cox}, {Danielson}, {Dannerbauer}, {Greve}, {Ivison}, {Karim}, \&
  {Thomson}}]{2015ApJ...806..110D}
{da Cunha}, E., {Walter}, F., {Smail}, I.~R., {et~al.} 2015, \apj, 806, 110

\bibitem[{{Daddi} {et~al.}(2007){Daddi}, {Dickinson}, {Morrison}, {Chary},
  {Cimatti}, {Elbaz}, {Frayer}, {Renzini}, {Pope}, {Alexander}, {Bauer},
  {Giavalisco}, {Huynh}, {Kurk}, \& {Mignoli}}]{2007ApJ...670..156D}
{Daddi}, E., {Dickinson}, M., {Morrison}, G., {et~al.} 2007, \apj, 670, 156

\bibitem[{{Dekel} {et~al.}(2009){Dekel}, {Birnboim}, {Engel}, {Freundlich},
  {Goerdt}, {Mumcuoglu}, {Neistein}, {Pichon}, {Teyssier}, \&
  {Zinger}}]{2009Natur.457..451D}
{Dekel}, A., {Birnboim}, Y., {Engel}, G., {et~al.} 2009, \nat, 457, 451

\bibitem[{{Dubois} {et~al.}(2015){Dubois}, {Volonteri}, {Silk}, {Devriendt},
  {Slyz}, \& {Teyssier}}]{2015MNRAS.452.1502D}
{Dubois}, Y., {Volonteri}, M., {Silk}, J., {et~al.} 2015, \mnras, 452, 1502

\bibitem[{{Elbaz} {et~al.}(2007){Elbaz}, {Daddi}, {Le Borgne}, {Dickinson},
  {Alexander}, {Chary}, {Starck}, {Brandt}, {Kitzbichler}, {MacDonald},
  {Nonino}, {Popesso}, {Stern}, \& {Vanzella}}]{2007A&A...468...33E}
{Elbaz}, D., {Daddi}, E., {Le Borgne}, D., {et~al.} 2007, \aap, 468, 33

\bibitem[{{Elbaz} {et~al.}(2018){Elbaz}, {Leiton}, {Nagar}, {Okumura},
  {Franco}, {Schreiber}, {Pannella}, {Wang}, {Dickinson}, {D{\'{\i}}az-Santos},
  {Ciesla}, {Daddi}, {Bournaud}, {Magdis}, {Zhou}, \&
  {Rujopakarn}}]{2018A&A...616A.110E}
{Elbaz}, D., {Leiton}, R., {Nagar}, N., {et~al.} 2018, \aap, 616, A110

\bibitem[{{Fan} {et~al.}(2014){Fan}, {Fang}, {Chen}, {Li}, {Lv}, {Knudsen}, \&
  {Kong}}]{2014ApJ...784L...9F}
{Fan}, L., {Fang}, G., {Chen}, Y., {et~al.} 2014, \apjl, 784, L9

\bibitem[{{Freundlich} {et~al.}(2019){Freundlich}, {Combes}, {Tacconi},
  {Genzel}, {Garcia-Burillo}, {Neri}, {Contini}, {Bolatto}, {Lilly},
  {Salom{\'e}}, {Bicalho}, {Boissier}, {Boone}, {Bouch{\'e}}, {Bournaud},
  {Burkert}, {Carollo}, {Cooper}, {Cox}, {Feruglio}, {F{\"o}rster Schreiber},
  {Juneau}, {Lippa}, {Lutz}, {Naab}, {Renzini}, {Saintonge}, {Sternberg},
  {Walter}, {Weiner}, {Wei{\ss}}, \& {Wuyts}}]{2019A&A...622A.105F}
{Freundlich}, J., {Combes}, F., {Tacconi}, L.~J., {et~al.} 2019, \aap, 622,
  A105

\bibitem[{{Fujimoto} {et~al.}(2017){Fujimoto}, {Ouchi}, {Shibuya}, \&
  {Nagai}}]{2017ApJ...850...83F}
{Fujimoto}, S., {Ouchi}, M., {Shibuya}, T., \& {Nagai}, H. 2017, \apj, 850, 83

\bibitem[{{Fujimoto} {et~al.}(2018){Fujimoto}, {Ouchi}, {Kohno}, {Yamaguchi},
  {Hatsukade}, {Ueda}, {Shibuya}, {Inoue}, {Oogi}, {Toft},
  {G{\'o}mez-Guijarro}, {Wang}, {Espada}, {Nagao}, {Tanaka}, {Ao}, {Umehata},
  {Taniguchi}, {Nakanishi}, {Rujopakarn}, {Ivison}, {Wang}, {Lee}, {Tadaki},
  {Tamura}, \& {Dunlop}}]{2018ApJ...861....7F}
{Fujimoto}, S., {Ouchi}, M., {Kohno}, K., {et~al.} 2018, \apj, 861, 7

\bibitem[{{Gabor} {et~al.}(2009){Gabor}, {Impey}, {Jahnke}, {Simmons}, {Trump},
  {Koekemoer}, {Brusa}, {Cappelluti}, {Schinnerer}, {Smol{\v c}i{\'c}},
  {Salvato}, {Rhodes}, {Mobasher}, {Capak}, {Massey}, {Leauthaud}, \&
  {Scoville}}]{2009ApJ...691..705G}
{Gabor}, J.~M., {Impey}, C.~D., {Jahnke}, K., {et~al.} 2009, \apj, 691, 705

\bibitem[{{Gonz{\'a}lez-L{\'o}pez} {et~al.}(2017){Gonz{\'a}lez-L{\'o}pez},
  {Bauer}, {Romero-Ca{\~n}izales}, {Kneissl}, {Villard}, {Carvajal}, {Kim},
  {Laporte}, {Anguita}, {Aravena}, {Bouwens}, {Bradley}, {Carrasco}, {Demarco},
  {Ford}, {Ibar}, {Infante}, {Messias}, {Mu{\~n}oz Arancibia}, {Nagar},
  {Padilla}, {Treister}, {Troncoso}, \& {Zitrin}}]{2017A&A...597A..41G}
{Gonz{\'a}lez-L{\'o}pez}, J., {Bauer}, F.~E., {Romero-Ca{\~n}izales}, C.,
  {et~al.} 2017, \aap, 597, A41

\bibitem[{{Grogin} {et~al.}(2011){Grogin}, {Kocevski}, {Faber}, {Ferguson},
  {Koekemoer}, {Riess}, {Acquaviva}, {Alexander}, {Almaini}, {Ashby}, {Barden},
  {Bell}, {Bournaud}, {Brown}, {Caputi}, {Casertano}, {Cassata}, {Castellano},
  {Challis}, {Chary}, {Cheung}, {Cirasuolo}, {Conselice}, {Roshan Cooray},
  {Croton}, {Daddi}, {Dahlen}, {Dav{\'e}}, {de Mello}, {Dekel}, {Dickinson},
  {Dolch}, {Donley}, {Dunlop}, {Dutton}, {Elbaz}, {Fazio}, {Filippenko},
  {Finkelstein}, {Fontana}, {Gardner}, {Garnavich}, {Gawiser}, {Giavalisco},
  {Grazian}, {Guo}, {Hathi}, {H{\"a}ussler}, {Hopkins}, {Huang}, {Huang},
  {Jha}, {Kartaltepe}, {Kirshner}, {Koo}, {Lai}, {Lee}, {Li}, {Lotz}, {Lucas},
  {Madau}, {McCarthy}, {McGrath}, {McIntosh}, {McLure}, {Mobasher},
  {Moustakas}, {Mozena}, {Nandra}, {Newman}, {Niemi}, {Noeske}, {Papovich},
  {Pentericci}, {Pope}, {Primack}, {Rajan}, {Ravindranath}, {Reddy}, {Renzini},
  {Rix}, {Robaina}, {Rodney}, {Rosario}, {Rosati}, {Salimbeni}, {Scarlata},
  {Siana}, {Simard}, {Smidt}, {Somerville}, {Spinrad}, {Straughn}, {Strolger},
  {Telford}, {Teplitz}, {Trump}, {van der Wel}, {Villforth}, {Wechsler},
  {Weiner}, {Wiklind}, {Wild}, {Wilson}, {Wuyts}, {Yan}, \&
  {Yun}}]{2011ApJS..197...35G}
{Grogin}, N.~A., {Kocevski}, D.~D., {Faber}, S.~M., {et~al.} 2011, \apjs, 197,
  35

\bibitem[{{Harrison} {et~al.}(2012){Harrison}, {Alexander}, {Mullaney},
  {Altieri}, {Coia}, {Charmandaris}, {Daddi}, {Dannerbauer}, {Dasyra}, {Del
  Moro}, {Dickinson}, {Hickox}, {Ivison}, {Kartaltepe}, {Le Floc'h}, {Leiton},
  {Magnelli}, {Popesso}, {Rovilos}, {Rosario}, \&
  {Swinbank}}]{2012ApJ...760L..15H}
{Harrison}, C.~M., {Alexander}, D.~M., {Mullaney}, J.~R., {et~al.} 2012, \apjl,
  760, L15

\bibitem[{{Hodge} {et~al.}(2016){Hodge}, {Swinbank}, {Simpson}, {Smail},
  {Walter}, {Alexander}, {Bertoldi}, {Biggs}, {Brandt}, {Chapman}, {Chen},
  {Coppin}, {Cox}, {Dannerbauer}, {Edge}, {Greve}, {Ivison}, {Karim},
  {Knudsen}, {Menten}, {Rix}, {Schinnerer}, {Wardlow}, {Weiss}, \& {van der
  Werf}}]{2016ApJ...833..103H}
{Hodge}, J.~A., {Swinbank}, A.~M., {Simpson}, J.~M., {et~al.} 2016, \apj, 833,
  103

\bibitem[{{Ikarashi} {et~al.}(2015){Ikarashi}, {Ivison}, {Caputi}, {Aretxaga},
  {Dunlop}, {Hatsukade}, {Hughes}, {Iono}, {Izumi}, {Kawabe}, {Kohno}, {Lagos},
  {Motohara}, {Nakanishi}, {Ohta}, {Tamura}, {Umehata}, {Wilson}, {Yabe}, \&
  {Yun}}]{2015ApJ...810..133I}
{Ikarashi}, S., {Ivison}, R.~J., {Caputi}, K.~I., {et~al.} 2015, \apj, 810, 133

\bibitem[{{Ilbert} {et~al.}(2015){Ilbert}, {Arnouts}, {Le Floc'h}, {Aussel},
  {Bethermin}, {Capak}, {Hsieh}, {Kajisawa}, {Karim}, {Le F{\`e}vre}, {Lee},
  {Lilly}, {McCracken}, {Michel-Dansac}, {Moutard}, {Renzini}, {Salvato},
  {Sanders}, {Scoville}, {Sheth}, {Silverman}, {Smol{\v c}i{\'c}}, {Taniguchi},
  \& {Tresse}}]{2015A&A...579A...2I}
{Ilbert}, O., {Arnouts}, S., {Le Floc'h}, E., {et~al.} 2015, \aap, 579, A2

\bibitem[{{Juneau} {et~al.}(2013){Juneau}, {Dickinson}, {Bournaud},
  {Alexander}, {Daddi}, {Mullaney}, {Magnelli}, {Kartaltepe}, {Hwang},
  {Willner}, {Coil}, {Rosario}, {Trump}, {Weiner}, {Willmer}, {Cooper},
  {Elbaz}, {Faber}, {Frayer}, {Kocevski}, {Laird}, {Monkiewicz}, {Nandra},
  {Newman}, {Salim}, \& {Symeonidis}}]{2013ApJ...764..176J}
{Juneau}, S., {Dickinson}, M., {Bournaud}, F., {et~al.} 2013, \apj, 764, 176

\bibitem[{{Kartaltepe} {et~al.}(2012){Kartaltepe}, {Dickinson}, {Alexander},
  {Bell}, {Dahlen}, {Elbaz}, {Faber}, {Lotz}, {McIntosh}, {Wiklind}, {Altieri},
  {Aussel}, {Bethermin}, {Bournaud}, {Charmandaris}, {Conselice}, {Cooray},
  {Dannerbauer}, {Dav{\'e}}, {Dunlop}, {Dekel}, {Ferguson}, {Grogin}, {Hwang},
  {Ivison}, {Kocevski}, {Koekemoer}, {Koo}, {Lai}, {Leiton}, {Lucas}, {Lutz},
  {Magdis}, {Magnelli}, {Morrison}, {Mozena}, {Mullaney}, {Newman}, {Pope},
  {Popesso}, {van der Wel}, {Weiner}, \& {Wuyts}}]{2012ApJ...757...23K}
{Kartaltepe}, J.~S., {Dickinson}, M., {Alexander}, D.~M., {et~al.} 2012, \apj,
  757, 23

\bibitem[{{Kennicutt} \& {Evans}(2012)}]{2012ARA&A..50..531K}
{Kennicutt}, R.~C., \& {Evans}, N.~J. 2012, \araa, 50, 531

\bibitem[{{Kennicutt}(1998)}]{1998ApJ...498..541K}
{Kennicutt}, Jr., R.~C. 1998, \apj, 498, 541

\bibitem[{{Kocevski} {et~al.}(2012){Kocevski}, {Faber}, {Mozena}, {Koekemoer},
  {Nandra}, {Rangel}, {Laird}, {Brusa}, {Wuyts}, {Trump}, {Koo}, {Somerville},
  {Bell}, {Lotz}, {Alexander}, {Bournaud}, {Conselice}, {Dahlen}, {Dekel},
  {Donley}, {Dunlop}, {Finoguenov}, {Georgakakis}, {Giavalisco}, {Guo},
  {Grogin}, {Hathi}, {Juneau}, {Kartaltepe}, {Lucas}, {McGrath}, {McIntosh},
  {Mobasher}, {Robaina}, {Rosario}, {Straughn}, {van der Wel}, \&
  {Villforth}}]{2012ApJ...744..148K}
{Kocevski}, D.~D., {Faber}, S.~M., {Mozena}, M., {et~al.} 2012, \apj, 744, 148

\bibitem[{{Koekemoer} {et~al.}(2011){Koekemoer}, {Faber}, {Ferguson}, {Grogin},
  {Kocevski}, {Koo}, {Lai}, {Lotz}, {Lucas}, {McGrath}, {Ogaz}, {Rajan},
  {Riess}, {Rodney}, {Strolger}, {Casertano}, {Castellano}, {Dahlen},
  {Dickinson}, {Dolch}, {Fontana}, {Giavalisco}, {Grazian}, {Guo}, {Hathi},
  {Huang}, {van der Wel}, {Yan}, {Acquaviva}, {Alexander}, {Almaini}, {Ashby},
  {Barden}, {Bell}, {Bournaud}, {Brown}, {Caputi}, {Cassata}, {Challis},
  {Chary}, {Cheung}, {Cirasuolo}, {Conselice}, {Roshan Cooray}, {Croton},
  {Daddi}, {Dav{\'e}}, {de Mello}, {de Ravel}, {Dekel}, {Donley}, {Dunlop},
  {Dutton}, {Elbaz}, {Fazio}, {Filippenko}, {Finkelstein}, {Frazer}, {Gardner},
  {Garnavich}, {Gawiser}, {Gruetzbauch}, {Hartley}, {H{\"a}ussler},
  {Herrington}, {Hopkins}, {Huang}, {Jha}, {Johnson}, {Kartaltepe},
  {Khostovan}, {Kirshner}, {Lani}, {Lee}, {Li}, {Madau}, {McCarthy},
  {McIntosh}, {McLure}, {McPartland}, {Mobasher}, {Moreira}, {Mortlock},
  {Moustakas}, {Mozena}, {Nandra}, {Newman}, {Nielsen}, {Niemi}, {Noeske},
  {Papovich}, {Pentericci}, {Pope}, {Primack}, {Ravindranath}, {Reddy},
  {Renzini}, {Rix}, {Robaina}, {Rosario}, {Rosati}, {Salimbeni}, {Scarlata},
  {Siana}, {Simard}, {Smidt}, {Snyder}, {Somerville}, {Spinrad}, {Straughn},
  {Telford}, {Teplitz}, {Trump}, {Vargas}, {Villforth}, {Wagner}, {Wandro},
  {Wechsler}, {Weiner}, {Wiklind}, {Wild}, {Wilson}, {Wuyts}, \&
  {Yun}}]{2011ApJS..197...36K}
{Koekemoer}, A.~M., {Faber}, S.~M., {Ferguson}, H.~C., {et~al.} 2011, \apjs,
  197, 36

\bibitem[{{Laigle} {et~al.}(2016){Laigle}, {McCracken}, {Ilbert}, {Hsieh},
  {Davidzon}, {Capak}, {Hasinger}, {Silverman}, {Pichon}, {Coupon}, {Aussel},
  {Le Borgne}, {Caputi}, {Cassata}, {Chang}, {Civano}, {Dunlop}, {Fynbo},
  {Kartaltepe}, {Koekemoer}, {Le F{\'e}vre}, {Le Floc'h}, {Leauthaud}, {Lilly},
  {Lin}, {Marchesi}, {Milvang-Jensen}, {Salvato}, {Sanders}, {Scoville},
  {Smolcic}, {Stockmann}, {Taniguchi}, {Tasca}, {Toft}, {Vaccari}, \&
  {Zabl}}]{2016ApJS..224...24L}
{Laigle}, C., {McCracken}, H.~J., {Ilbert}, O., {et~al.} 2016, \apjs, 224, 24

\bibitem[{{Lang} {et~al.}(2019){Lang}, {Schinnerer}, {Smail},
  {Dudzevi{\v{c}}i{\={u}}t{\.{e}}}, {Swinbank}, {Liu}, {Leslie}, {Almaini},
  {An}, {Bertoldi}, {Blain}, {Chapman}, {Chen}, {Conselice}, {Cooke}, {Coppin},
  {Dunlop}, {Farrah}, {Fudamoto}, {Geach}, {Gullberg}, {Harrington}, {Hodge},
  {Ivison}, {Jim{\'e}nez-Andrade}, {Magnelli}, {Micha{\l}owski}, {Oesch},
  {Scott}, {Simpson}, {Smol{\v{c}}i{\'c}}, {Stach}, {Thomson}, {Toft},
  {Vardoulaki}, {Wardlow}, {Weiss}, \& {van der Werf}}]{2019ApJ...879...54L}
{Lang}, P., {Schinnerer}, E., {Smail}, I., {et~al.} 2019, \apj, 879, 54

\bibitem[{{Lindroos} {et~al.}(2015){Lindroos}, {Knudsen}, {Vlemmings},
  {Conway}, \& {Mart{\'{\i}}-Vidal}}]{2015MNRAS.446.3502L}
{Lindroos}, L., {Knudsen}, K.~K., {Vlemmings}, W., {Conway}, J., \&
  {Mart{\'{\i}}-Vidal}, I. 2015, \mnras, 446, 3502

\bibitem[{{Mainieri} {et~al.}(2011){Mainieri}, {Bongiorno}, {Merloni}, {Aller},
  {Carollo}, {Iwasawa}, {Koekemoer}, {Mignoli}, {Silverman}, {Bolzonella},
  {Brusa}, {Comastri}, {Gilli}, {Halliday}, {Ilbert}, {Lusso}, {Salvato},
  {Vignali}, {Zamorani}, {Contini}, {Kneib}, {Le F{\`e}vre}, {Lilly},
  {Renzini}, {Scodeggio}, {Balestra}, {Bardelli}, {Caputi}, {Coppa},
  {Cucciati}, {de la Torre}, {de Ravel}, {Franzetti}, {Garilli}, {Iovino},
  {Kampczyk}, {Knobel}, {Kova{\v c}}, {Lamareille}, {Le Borgne}, {Le Brun},
  {Maier}, {Nair}, {Pello}, {Peng}, {Perez Montero}, {Pozzetti},
  {Ricciardelli}, {Tanaka}, {Tasca}, {Tresse}, {Vergani}, {Zucca}, {Aussel},
  {Capak}, {Cappelluti}, {Elvis}, {Fiore}, {Hasinger}, {Impey}, {Le Floc'h},
  {Scoville}, {Taniguchi}, \& {Trump}}]{2011A&A...535A..80M}
{Mainieri}, V., {Bongiorno}, A., {Merloni}, A., {et~al.} 2011, \aap, 535, A80

\bibitem[{{Marchesi} {et~al.}(2016){Marchesi}, {Civano}, {Elvis}, {Salvato},
  {Brusa}, {Comastri}, {Gilli}, {Hasinger}, {Lanzuisi}, {Miyaji}, {Treister},
  {Urry}, {Vignali}, {Zamorani}, {Allevato}, {Cappelluti}, {Cardamone},
  {Finoguenov}, {Griffiths}, {Karim}, {Laigle}, {LaMassa}, {Jahnke}, {Ranalli},
  {Schawinski}, {Schinnerer}, {Silverman}, {Smolcic}, {Suh}, \&
  {Trakhtenbrot}}]{2016ApJ...817...34M}
{Marchesi}, S., {Civano}, F., {Elvis}, M., {et~al.} 2016, \apj, 817, 34

\bibitem[{{McMullin} {et~al.}(2007){McMullin}, {Waters}, {Schiebel}, {Young},
  \& {Golap}}]{2007ASPC..376..127M}
{McMullin}, J.~P., {Waters}, B., {Schiebel}, D., {Young}, W., \& {Golap}, K.
  2007, in Astronomical Society of the Pacific Conference Series, Vol. 376,
  Astronomical Data Analysis Software and Systems XVI, ed. R.~A. {Shaw},
  F.~{Hill}, \& D.~J. {Bell}, 127

\bibitem[{{Mechtley} {et~al.}(2016){Mechtley}, {Jahnke}, {Windhorst}, {Andrae},
  {Cisternas}, {Cohen}, {Hewlett}, {Koekemoer}, {Schramm}, {Schulze},
  {Silverman}, {Villforth}, {van der Wel}, \& {Wisotzki}}]{2016ApJ...830..156M}
{Mechtley}, M., {Jahnke}, K., {Windhorst}, R.~A., {et~al.} 2016, \apj, 830, 156

\bibitem[{{Mullaney} {et~al.}(2011){Mullaney}, {Alexander}, {Goulding}, \&
  {Hickox}}]{2011MNRAS.414.1082M}
{Mullaney}, J.~R., {Alexander}, D.~M., {Goulding}, A.~D., \& {Hickox}, R.~C.
  2011, \mnras, 414, 1082

\bibitem[{{Mullaney} {et~al.}(2012){Mullaney}, {Pannella}, {Daddi},
  {Alexander}, {Elbaz}, {Hickox}, {Bournaud}, {Altieri}, {Aussel}, {Coia},
  {Dannerbauer}, {Dasyra}, {Dickinson}, {Hwang}, {Kartaltepe}, {Leiton},
  {Magdis}, {Magnelli}, {Popesso}, {Valtchanov}, {Bauer}, {Brandt}, {Del Moro},
  {Hanish}, {Ivison}, {Juneau}, {Luo}, {Lutz}, {Sargent}, {Scott}, \&
  {Xue}}]{2012MNRAS.419...95M}
{Mullaney}, J.~R., {Pannella}, M., {Daddi}, E., {et~al.} 2012, \mnras, 419, 95

\bibitem[{{Noeske} {et~al.}(2007){Noeske}, {Weiner}, {Faber}, {Papovich},
  {Koo}, {Somerville}, {Bundy}, {Conselice}, {Newman}, {Schiminovich}, {Le
  Floc'h}, {Coil}, {Rieke}, {Lotz}, {Primack}, {Barmby}, {Cooper}, {Davis},
  {Ellis}, {Fazio}, {Guhathakurta}, {Huang}, {Kassin}, {Martin}, {Phillips},
  {Rich}, {Small}, {Willmer}, \& {Wilson}}]{2007ApJ...660L..43N}
{Noeske}, K.~G., {Weiner}, B.~J., {Faber}, S.~M., {et~al.} 2007, \apjl, 660,
  L43

\bibitem[{{Peng} {et~al.}(2010){Peng}, {Ho}, {Impey}, \&
  {Rix}}]{2010AJ....139.2097P}
{Peng}, C.~Y., {Ho}, L.~C., {Impey}, C.~D., \& {Rix}, H.-W. 2010, \aj, 139,
  2097

\bibitem[{{Polletta} {et~al.}(2007){Polletta}, {Tajer}, {Maraschi},
  {Trinchieri}, {Lonsdale}, {Chiappetti}, {Andreon}, {Pierre}, {Le F{\`e}vre},
  {Zamorani}, {Maccagni}, {Garcet}, {Surdej}, {Franceschini}, {Alloin},
  {Shupe}, {Surace}, {Fang}, {Rowan-Robinson}, {Smith}, \&
  {Tresse}}]{2007ApJ...663...81P}
{Polletta}, M., {Tajer}, M., {Maraschi}, L., {et~al.} 2007, \apj, 663, 81

\bibitem[{{Prieto} {et~al.}(2010){Prieto}, {Reunanen}, {Tristram}, {Neumayer},
  {Fernandez-Ontiveros}, {Orienti}, \& {Meisenheimer}}]{2010MNRAS.402..724P}
{Prieto}, M.~A., {Reunanen}, J., {Tristram}, K.~R.~W., {et~al.} 2010, \mnras,
  402, 724

\bibitem[{{Richards} {et~al.}(2006){Richards}, {Lacy}, {Storrie-Lombardi},
  {Hall}, {Gallagher}, {Hines}, {Fan}, {Papovich}, {Vanden Berk}, {Trammell},
  {Schneider}, {Vestergaard}, {York}, {Jester}, {Anderson}, {Budav{\'a}ri}, \&
  {Szalay}}]{2006ApJS..166..470R}
{Richards}, G.~T., {Lacy}, M., {Storrie-Lombardi}, L.~J., {et~al.} 2006, \apjs,
  166, 470

\bibitem[{{Rosario} {et~al.}(2013){Rosario}, {Santini}, {Lutz}, {Netzer},
  {Bauer}, {Berta}, {Magnelli}, {Popesso}, {Alexander}, {Brandt}, {Genzel},
  {Maiolino}, {Mullaney}, {Nordon}, {Saintonge}, {Tacconi}, \&
  {Wuyts}}]{2013ApJ...771...63R}
{Rosario}, D.~J., {Santini}, P., {Lutz}, D., {et~al.} 2013, \apj, 771, 63

\bibitem[{{Rujopakarn} {et~al.}(2016){Rujopakarn}, {Dunlop}, {Rieke}, {Ivison},
  {Cibinel}, {Nyland}, {Jagannathan}, {Silverman}, {Alexander}, {Biggs},
  {Bhatnagar}, {Ballantyne}, {Dickinson}, {Elbaz}, {Geach}, {Hayward},
  {Kirkpatrick}, {McLure}, {Micha{\l}owski}, {Miller}, {Narayanan}, {Owen},
  {Pannella}, {Papovich}, {Pope}, {Rau}, {Robertson}, {Scott}, {Swinbank}, {van
  der Werf}, {van Kampen}, {Weiner}, \& {Windhorst}}]{2016ApJ...833...12R}
{Rujopakarn}, W., {Dunlop}, J.~S., {Rieke}, G.~H., {et~al.} 2016, \apj, 833, 12

\bibitem[{{Salvato} {et~al.}(2011){Salvato}, {Ilbert}, {Hasinger}, {Rau},
  {Civano}, {Zamorani}, {Brusa}, {Elvis}, {Vignali}, {Aussel}, {Comastri},
  {Fiore}, {Le Floc'h}, {Mainieri}, {Bardelli}, {Bolzonella}, {Bongiorno},
  {Capak}, {Caputi}, {Cappelluti}, {Carollo}, {Contini}, {Garilli}, {Iovino},
  {Fotopoulou}, {Fruscione}, {Gilli}, {Halliday}, {Kneib}, {Kakazu},
  {Kartaltepe}, {Koekemoer}, {Kovac}, {Ideue}, {Ikeda}, {Impey}, {Le Fevre},
  {Lamareille}, {Lanzuisi}, {Le Borgne}, {Le Brun}, {Lilly}, {Maier},
  {Manohar}, {Masters}, {McCracken}, {Messias}, {Mignoli}, {Mobasher}, {Nagao},
  {Pello}, {Puccetti}, {Perez-Montero}, {Renzini}, {Sargent}, {Sanders},
  {Scodeggio}, {Scoville}, {Shopbell}, {Silvermann}, {Taniguchi}, {Tasca},
  {Tresse}, {Trump}, \& {Zucca}}]{2011ApJ...742...61S}
{Salvato}, M., {Ilbert}, O., {Hasinger}, G., {et~al.} 2011, \apj, 742, 61

\bibitem[{{Schreiber} {et~al.}(2015){Schreiber}, {Pannella}, {Elbaz},
  {B{\'e}thermin}, {Inami}, {Dickinson}, {Magnelli}, {Wang}, {Aussel}, {Daddi},
  {Juneau}, {Shu}, {Sargent}, {Buat}, {Faber}, {Ferguson}, {Giavalisco},
  {Koekemoer}, {Magdis}, {Morrison}, {Papovich}, {Santini}, \&
  {Scott}}]{2015A&A...575A..74S}
{Schreiber}, C., {Pannella}, M., {Elbaz}, D., {et~al.} 2015, \aap, 575, A74

\bibitem[{{Scoville} {et~al.}(2007){Scoville}, {Aussel}, {Brusa}, {Capak},
  {Carollo}, {Elvis}, {Giavalisco}, {Guzzo}, {Hasinger}, {Impey}, {Kneib},
  {LeFevre}, {Lilly}, {Mobasher}, {Renzini}, {Rich}, {Sanders}, {Schinnerer},
  {Schminovich}, {Shopbell}, {Taniguchi}, \& {Tyson}}]{2007ApJS..172....1S}
{Scoville}, N., {Aussel}, H., {Brusa}, M., {et~al.} 2007, \apjs, 172, 1

\bibitem[{{Silverman} {et~al.}(2011){Silverman}, {Kampczyk}, {Jahnke},
  {Andrae}, {Lilly}, {Elvis}, {Civano}, {Mainieri}, {Vignali}, {Zamorani},
  {Nair}, {Le F{\`e}vre}, {de Ravel}, {Bardelli}, {Bongiorno}, {Bolzonella},
  {Cappi}, {Caputi}, {Carollo}, {Contini}, {Coppa}, {Cucciati}, {de la Torre},
  {Franzetti}, {Garilli}, {Halliday}, {Hasinger}, {Iovino}, {Knobel},
  {Koekemoer}, {Kova{\v c}}, {Lamareille}, {Le Borgne}, {Le Brun}, {Maier},
  {Mignoli}, {Pello}, {P{\'e}rez-Montero}, {Ricciardelli}, {Peng}, {Scodeggio},
  {Tanaka}, {Tasca}, {Tresse}, {Vergani}, {Zucca}, {Brusa}, {Cappelluti},
  {Comastri}, {Finoguenov}, {Fu}, {Gilli}, {Hao}, {Ho}, \&
  {Salvato}}]{2011ApJ...743....2S}
{Silverman}, J.~D., {Kampczyk}, P., {Jahnke}, K., {et~al.} 2011, \apj, 743, 2

\bibitem[{{Simpson} {et~al.}(2015){Simpson}, {Smail}, {Swinbank}, {Chapman},
  {Geach}, {Ivison}, {Thomson}, {Aretxaga}, {Blain}, {Cowley}, {Chen},
  {Coppin}, {Dunlop}, {Edge}, {Farrah}, {Ibar}, {Karim}, {Knudsen},
  {Meijerink}, {Micha{\l}owski}, {Scott}, {Spaans}, \& {van der
  Werf}}]{2015ApJ...807..128S}
{Simpson}, J.~M., {Smail}, I., {Swinbank}, A.~M., {et~al.} 2015, \apj, 807, 128

\bibitem[{{Simpson} {et~al.}(2017){Simpson}, {Smail}, {Wang}, {Riechers},
  {Dunlop}, {Ao}, {Bourne}, {Bunker}, {Chapman}, {Chen}, {Dannerbauer},
  {Geach}, {Goto}, {Harrison}, {Hwang}, {Ivison}, {Kodama}, {Lee}, {Lee},
  {Lee}, {Lim}, {Micha{\l}owski}, {Rosario}, {Shim}, {Shu}, {Swinbank}, {Tee},
  {Toba}, {Valiante}, {Wang}, \& {Zheng}}]{2017ApJ...844L..10S}
{Simpson}, J.~M., {Smail}, I., {Wang}, W.-H., {et~al.} 2017, \apjl, 844, L10

\bibitem[{{Speagle} {et~al.}(2014){Speagle}, {Steinhardt}, {Capak}, \&
  {Silverman}}]{2014ApJS..214...15S}
{Speagle}, J.~S., {Steinhardt}, C.~L., {Capak}, P.~L., \& {Silverman}, J.~D.
  2014, \apjs, 214, 15

\bibitem[{{Tacchella} {et~al.}(2016){Tacchella}, {Dekel}, {Carollo},
  {Ceverino}, {DeGraf}, {Lapiner}, {Mandelker}, \& {Primack
  Joel}}]{2016MNRAS.457.2790T}
{Tacchella}, S., {Dekel}, A., {Carollo}, C.~M., {et~al.} 2016, \mnras, 457,
  2790

\bibitem[{{Tacconi} {et~al.}(2018){Tacconi}, {Genzel}, {Saintonge}, {Combes},
  {Garc{\'{\i}}a-Burillo}, {Neri}, {Bolatto}, {Contini}, {F{\"o}rster
  Schreiber}, {Lilly}, {Lutz}, {Wuyts}, {Accurso}, {Boissier}, {Boone},
  {Bouch{\'e}}, {Bournaud}, {Burkert}, {Carollo}, {Cooper}, {Cox}, {Feruglio},
  {Freundlich}, {Herrera-Camus}, {Juneau}, {Lippa}, {Naab}, {Renzini},
  {Salome}, {Sternberg}, {Tadaki}, {{\"U}bler}, {Walter}, {Weiner}, \&
  {Weiss}}]{2018ApJ...853..179T}
{Tacconi}, L.~J., {Genzel}, R., {Saintonge}, A., {et~al.} 2018, \apj, 853, 179

\bibitem[{{Tadaki} {et~al.}(2017){Tadaki}, {Kodama}, {Nelson}, {Belli},
  {F{\"o}rster Schreiber}, {Genzel}, {Hayashi}, {Herrera-Camus}, {Koyama},
  {Lang}, {Lutz}, {Shimakawa}, {Tacconi}, {{\"U}bler}, {Wisnioski}, {Wuyts},
  {Hatsukade}, {Lippa}, {Nakanishi}, {Ikarashi}, {Kohno}, {Suzuki}, {Tamura},
  \& {Tanaka}}]{2017ApJ...841L..25T}
{Tadaki}, K.-i., {Kodama}, T., {Nelson}, E.~J., {et~al.} 2017, \apjl, 841, L25

\bibitem[{{Treister} {et~al.}(2009){Treister}, {Urry}, \&
  {Virani}}]{2009ApJ...696..110T}
{Treister}, E., {Urry}, C.~M., \& {Virani}, S. 2009, \apj, 696, 110

\bibitem[{{van der Wel} {et~al.}(2012){van der Wel}, {Bell}, {H{\"a}ussler},
  {McGrath}, {Chang}, {Guo}, {McIntosh}, {Rix}, {Barden}, {Cheung}, {Faber},
  {Ferguson}, {Galametz}, {Grogin}, {Hartley}, {Kartaltepe}, {Kocevski},
  {Koekemoer}, {Lotz}, {Mozena}, {Peth}, \& {Peng}}]{2012ApJS..203...24V}
{van der Wel}, A., {Bell}, E.~F., {H{\"a}ussler}, B., {et~al.} 2012, \apjs,
  203, 24

\bibitem[{{van der Wel} {et~al.}(2014){van der Wel}, {Franx}, {van Dokkum},
  {Skelton}, {Momcheva}, {Whitaker}, {Brammer}, {Bell}, {Rix}, {Wuyts},
  {Ferguson}, {Holden}, {Barro}, {Koekemoer}, {Chang}, {McGrath},
  {H{\"a}ussler}, {Dekel}, {Behroozi}, {Fumagalli}, {Leja}, {Lundgren},
  {Maseda}, {Nelson}, {Wake}, {Patel}, {Labb{\'e}}, {Faber}, {Grogin}, \&
  {Kocevski}}]{2014ApJ...788...28V}
{van der Wel}, A., {Franx}, M., {van Dokkum}, P.~G., {et~al.} 2014, \apj, 788,
  28

\bibitem[{{Villforth} {et~al.}(2014){Villforth}, {Hamann}, {Rosario},
  {Santini}, {McGrath}, {van der Wel}, {Chang}, {Guo}, {Dahlen}, {Bell},
  {Conselice}, {Croton}, {Dekel}, {Faber}, {Grogin}, {Hamilton}, {Hopkins},
  {Juneau}, {Kartaltepe}, {Kocevski}, {Koekemoer}, {Koo}, {Lotz}, {McIntosh},
  {Mozena}, {Somerville}, \& {Wild}}]{2014MNRAS.439.3342V}
{Villforth}, C., {Hamann}, F., {Rosario}, D.~J., {et~al.} 2014, \mnras, 439,
  3342

\bibitem[{{Villforth} {et~al.}(2017){Villforth}, {Hamilton}, {Pawlik},
  {Hewlett}, {Rowlands}, {Herbst}, {Shankar}, {Fontana}, {Hamann}, {Koekemoer},
  {Pforr}, {Trump}, \& {Wuyts}}]{2017MNRAS.466..812V}
{Villforth}, C., {Hamilton}, T., {Pawlik}, M.~M., {et~al.} 2017, \mnras, 466,
  812

\bibitem[{{Whitaker} {et~al.}(2014){Whitaker}, {Franx}, {Leja}, {van Dokkum},
  {Henry}, {Skelton}, {Fumagalli}, {Momcheva}, {Brammer}, {Labb{\'e}},
  {Nelson}, \& {Rigby}}]{2014ApJ...795..104W}
{Whitaker}, K.~E., {Franx}, M., {Leja}, J., {et~al.} 2014, \apj, 795, 104

\bibitem[{{Williams} {et~al.}(2009){Williams}, {Quadri}, {Franx}, {van Dokkum},
  \& {Labb{\'e}}}]{2009ApJ...691.1879W}
{Williams}, R.~J., {Quadri}, R.~F., {Franx}, M., {van Dokkum}, P., \&
  {Labb{\'e}}, I. 2009, \apj, 691, 1879

\bibitem[{{Wuyts} {et~al.}(2012){Wuyts}, {F{\"o}rster Schreiber}, {Genzel},
  {Guo}, {Barro}, {Bell}, {Dekel}, {Faber}, {Ferguson}, {Giavalisco}, {Grogin},
  {Hathi}, {Huang}, {Kocevski}, {Koekemoer}, {Koo}, {Lotz}, {Lutz}, {McGrath},
  {Newman}, {Rosario}, {Saintonge}, {Tacconi}, {Weiner}, \& {van der
  Wel}}]{2012ApJ...753..114W}
{Wuyts}, S., {F{\"o}rster Schreiber}, N.~M., {Genzel}, R., {et~al.} 2012, The
  Astrophysical Journal, 753, 114

\bibitem[{{Zolotov} {et~al.}(2015){Zolotov}, {Dekel}, {Mandelker}, {Tweed},
  {Inoue}, {DeGraf}, {Ceverino}, {Primack}, {Barro}, \&
  {Faber}}]{2015MNRAS.450.2327Z}
{Zolotov}, A., {Dekel}, A., {Mandelker}, N., {et~al.} 2015, \mnras, 450, 2327

\end{thebibliography}



\end{document}